# A CEERS Discovery of an Accreting Supermassive Black Hole 570 Myr after the Big Bang: Identifying a Progenitor of Massive z > 6 Quasars

Rebecca L. Larson[1,40], Steven L. Finkelstein[1], Dale D. Kocevski[2], Taylor A. Hutchison[3,41], Jonathan R. Trump[4],
Pablo Arrabal Haro[5], Volker Bromm[6], Nikko J. Cleri[7,8], Mark Dickinson[5], Seiji Fujimoto[1], Jeyhan S. Kartaltepe[9],
Anton M. Koekemoer[10], Casey Papovich[7,8], Nor Pirzkal[11], Sandro Tacchella[12,13], Jorge A. Zavala[14],
Micaela Bagley[1], Peter Behroozi[15,16], Jaclyn B. Champagne[17], Justin W. Cole[7,8], Intae Jung[10], Alexa M. Morales[18],
Guang Yang[19,20], Haowen Zhang[15], Adi Zitrin[21], Ricardo O. Amorín[22,23], Denis Burgarella[24], Caitlin M. Casey[25,26],
Óscar A. Chávez Ortiz[6], Isabella G. Cox[9], Katherine Chworowsky[6,40], Adriano Fontana[27], Eric Gawiser[28],
Andrea Grazian[29], Norman A. Grogin[10], Santosh Harish[9], Nimish P. Hathi[30], Michaela Hirschmann[31],
Benne W. Holwerda[32], Stéphanie Juneau[33], Gene C. K. Leung[34], Ray A. Lucas[10], Elizabeth J. McGrath[2],
Pablo G. Pérez-González[35], Jane R. Rigby[3], Lise-Marie Seillé[24], Raymond C. Simons[4], Alexander de la Vega[36],
Benjamin J. Weiner[37], Stephen M. Wilkins[38,39], and L. Y. Aaron Yung[3,41]
and The CEERS Team

[1] The University of Texas at Austin, Department of Astronomy, Austin, TX, USA; rllsps@rit.edu
[2] Department of Physics and Astronomy, Colby College, Waterville, ME 04901, USA
[3] Astrophysics Science Division, NASA Goddard Space Flight Center, 8800 Greenbelt Rd, Greenbelt, MD 20771, USA
[4] Department of Physics, 196 Auditorium Road, Unit 3046, University of Connecticut, Storrs, CT 06269, USA
[5] NSF's National Optical-Infrared Astronomy Research Laboratory, 950 N. Cherry Ave., Tucson, AZ 85719, USA
[6] Department of Astronomy, The University of Texas at Austin, Austin, TX, USA
[7] Department of Physics and Astronomy, Texas A&M University, College Station, TX 77843-4242 USA
[8] George P. and Cynthia Woods Mitchell Institute for Fundamental Physics and Astronomy, Texas A&M University, College Station, TX 77843-4242 USA
[9] Laboratory for Multiwavelength Astrophysics, School of Physics and Astronomy, Rochester Institute of Technology, 84 Lomb Memorial Drive, Rochester, NY 14623, USA
[10] Space Telescope Science Institute, 3700 San Martin Drive, Baltimore, MD 21218, USA
[11] ESA/AURA Space Telescope Science Institute, Baltimore, MD 21218, USA
[12] Kavli Institute for Cosmology, University of Cambridge, Madingley Road, Cambridge, CB3 0HA, UK
[13] Cavendish Laboratory, University of Cambridge, 19 JJ Thomson Avenue, Cambridge, CB3 0HE, UK
[14] National Astronomical Observatory of Japan, 2-21-1 Osawa, Mitaka, Tokyo 181-8588, Japan
[15] Department of Astronomy and Steward Observatory, University of Arizona, Tucson, AZ 85721, USA
[16] Division of Science, National Astronomical Observatory of Japan, 2-21-1 Osawa, Mitaka, Tokyo 181-8588, Japan
[17] Steward Observatory, University of Arizona, 933 N. Cherry Ave, Tucson, AZ 85719, USA
[18] Department of Astronomy, The University of Texas at Austin, 2515 Speedway, Austin, TX 78712, USA
[19] Kapteyn Astronomical Institute, University of Groningen, P.O. Box 800, 9700 AV Groningen, The Netherlands
[20] SRON Netherlands Institute for Space Research, Postbus 800, 9700 AV Groningen, The Netherlands
[21] Physics Department, Ben-Gurion University of the Negev, P.O. Box 653, Be'er-Sheva 84105, Israel
[22] Instituto de Investigación Multidisciplinar en Ciencia y Tecnología, Universidad de La Serena, Raul Bitrán 1305, La Serena 2204000, Chile
[23] Departamento de Astronomía, Universidad de La Serena, Av. Juan Cisternas 1200 Norte, La Serena 1720236, Chile
[24] Aix Marseille Univ, CNRS, CNES, LAM Marseille, France
[25] The University of Texas at Austin, 2515 Speedway Blvd Stop C1400, Austin, TX 78712, USA
[26] Cosmic Dawn Center (DAWN), Denmark
[27] INAF—Osservatorio Astronomico di Roma, via di Frascati 33, I-00078 Monte Porzio Catone, Italy
[28] Department of Physics and Astronomy, Rutgers, the State University of New Jersey, Piscataway, NJ 08854, USA
[29] INAF–Osservatorio Astronomico di Padova, Vicolo dell'Osservatorio 5, I-35122, Padova, Italy
[30] Space Telescope Science Institute, Baltimore, MD, USA
[31] Institute of Physics, Laboratory of Galaxy Evolution, Ecole Polytechnique Fdrale de Lausanne (EPFL), Observatoire de Sauverny, 1290 Versoix, Switzerland
[32] Physics & Astronomy Department, University of Louisville, Louisville, KY 40292, USA
[33] NSF's NOIRLab, 950 N. Cherry Ave., Tucson, AZ 85719, USA
[34] Department of Astronomy, The University of Texas at Austin USA
[35] Centro de Astrobiología (CAB), CSIC-INTA, Ctra. de Ajalvir km 4, Torrejón de Ardoz, E-28850, Madrid, Spain
[36] Department of Physics and Astronomy, University of California, 900 University Ave, Riverside, CA 92521, USA
[37] MMT/Steward Observatory, University of Arizona, 933 N. Cherry Ave., Tucson, AZ 85721, USA
[38] Astronomy Centre, University of Sussex, Falmer, Brighton, BN1 9QH, UK
[39] Institute of Space Sciences and Astronomy, University of Malta, Msida MSD 2080, Malta

## Abstract

We report the discovery of an accreting supermassive black hole at $z = 8.679$. This galaxy, denoted here as CEERS_1019, was previously discovered as a Lyα-break galaxy by Hubble with a Lyα redshift from Keck. As part of the Cosmic Evolution Early Release Science (CEERS) survey, we have observed this source with JWST/NIRSpec, MIRI, NIRCam, and NIRCam/WFSS and uncovered a plethora of emission lines. The H$\beta$ line is best fit by a narrow plus a broad component, where the latter is measured at 2.5$\sigma$ with an FWHM $\sim$1200 km s$^{-1}$. We conclude this originates in the broadline region of an active galactic nucleus (AGN). This is supported by the presence of weak high-ionization lines (N V, N IV], and C III]), as well as a spatial point-source component. The implied mass of the black hole (BH) is log ($M_{BH}/M_\odot$) = 6.95 $\pm$ 0.37, and we estimate that it is accreting at 1.2 $\pm$ 0.5 times the Eddington limit. The 1–8 $\mu$m photometric spectral energy distribution shows a continuum dominated by starlight and constrains the host galaxy to be massive (log M/M$_\odot$ $\sim$9.5) and highly star-forming (star formation rate, or SFR $\sim$ 30 M$_\odot$ yr$^{-1}$; log sSFR $\sim$ $-$ 7.9 yr$^{-1}$). The line ratios show that the gas is metal-poor (Z/Z$_\odot$ $\sim$ 0.1), dense ($n_e$ $\sim$ 10$^3$ cm$^{-3}$), and highly ionized (log U $\sim$ $-$ 2.1). We use this present highest-redshift AGN discovery to place constraints on BH seeding models and find that a combination of either super-Eddington accretion from stellar seeds or Eddington accretion from very massive BH seeds is required to form this object.

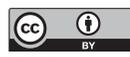

---

[40] NSF Graduate Fellow.
[41] NASA Postdoctoral Fellow.

*Unified Astronomy Thesaurus concepts:* AGN host galaxies (2017); Black holes (162); High-redshift galaxies (734); Galaxies (573); Infrared spectroscopy (2285); Spectroscopy (1558); Observational astronomy (1145)

## 1. Introduction

One of the most consequential periods in cosmic history is the Epoch of Reionization (EoR), where the material between galaxies underwent a significant transition from neutral to ionized hydrogen. Satisfactory explanations for the sources of radiation that contributed to this process, when it started, and how long it lasted are all presently wanting, primarily from the lack of necessary observatories and instruments tuned to the early Universe—until now. With the launch of JWST, we are on the cusp of an exciting new era in astronomy, as detailed studies of galaxies in the first billion years are finally possible.

One of the key questions JWST was designed to answer was when and how the first black holes (BHs) formed. Defining this epoch will help constrain the role these sources played in reionization alongside ionizing photons from massive stars. Supermassive black holes (SMBHs) exist at the centers of massive galaxies in the present-day Universe and exhibit a well-studied correlation with the velocity dispersions (and stellar masses) of galaxy bulge components (see the review by Kormendy & Ho 2013). These massive objects have >13 Gyr of cosmic time to grow to their present-day mass, which is possible via standard accretion scenarios with stellar-mass BH seeds ($\sim$1–10 $M_\odot$; possibly up to 100 M$_\odot$ if a Population III star remnant). However, the surprising discovery of $z > 6$ quasars with the Sloan Digital Sky Survey (SDSS) with BH masses of log ($M_{BH}/M_\odot$) > 9 (e.g., Fan et al. 2006; Bañados et al. 2018; Wang et al. 2021; Farina et al. 2022) challenges models of BH growth. Such objects require very early ($z \sim$ 25–30) stellar-mass seeds with near-unity duty cycles of Eddington-limited accretion and/or super-Eddington accretion to grow to such a mass from a stellar-mass seed in $\lesssim$ 1 Gyr (e.g., Volonteri et al. 2021 and references therein). None of these scenarios are adequately predicted by simulations (e.g., Volonteri et al. 2021; Fontanot et al. 2023). This tension is further exacerbated by the similarly massive quasars now being discovered at $z > 7$ (e.g., Mortlock et al. 2011; Bañados et al. 2018; see Fan et al. 2022 for a recent review), with the current highest-redshift known active galactic nucleus (AGN) being a bright quasar at $z$ = 7.64 (Wang et al. 2021).

These surprisingly massive BHs in the first Gyr of cosmic history have led to an alternative seeding theory: direct collapse black holes (DCBHs; Bromm & Loeb 2003). In this model, minihalos irradiated by Lyman–Werner photons (11.2–13.6 eV) cannot form molecular gas and thus do not form Population III stars. As these halos grow, they eventually cross the atomic cooling regime, at which point the gas will begin to cool rapidly via H I cooling (primarily Ly$\alpha$). Further fragmentation-inducing cooling is avoided due to the H$_2$-suppressing UV background, leading to collapse into massive BHs in the range of 10$^{4-6}$ $M_\odot$, preceded by a brief phase as a supermassive star (see Smith & Bromm 2019; Woods et al. 2019 and references therein for a detailed explanation of this process). A similarly massive BH seed could be formed as the remnant of a supermassive star powered by WIMP-like dark matter annihilation (so-called "dark stars"; Ilie et al. 2012; Freese et al. 2016). Such massive BH seeds, forming at $z \sim$ 10–15 (by necessity after the first generation of stars), could more feasibly grow into the observed $z \sim$ 6–7 quasar population (e.g., Madau et al. 2014; Natarajan et al. 2017; Regan et al. 2019; Latif et al. 2021; Pacucci & Loeb 2022; Massonneau et al. 2023; Trinca et al. 2023). These massive quasars must represent only the extreme cases—there likely exists a much larger population of lower-mass BHs and/or obscured BHs waiting to be discovered.

While DCBHs could alleviate the tension between observed BH masses and our theories of BH growth, such objects have yet to be observed. One clear pathway to understanding BH growth is to observe more SMBHs in the EoR. Identifying modest-sized BHs at earlier cosmic times could provide further evidence as to whether DCBHs are a necessary pathway. Additionally, the discovery of such a population would both better explain how the observed $z \sim$ 6 quasar population originally formed and inform on the potential contribution of AGNs to reionization both through X-ray heating (e.g., Jeon et al. 2014) and through ionizing photon contributions (e.g., Finkelstein et al. 2019; Giallongo et al. 2019; Grazian et al. 2020, 2022; Yung et al. 2021).

Prior to JWST, only the most massive SMBHs at high redshifts could be identified. However, the spectroscopic capabilities of JWST now enable the search for signs of AGN activity from less luminous sources, particularly those embedded in galaxies whose stellar emission dominates the total galaxy luminosity (e.g., Endsley et al. 2022; Kocevski et al. 2023) and/or where the bulk of the accretion emission is obscured (e.g., Fujimoto et al. 2022; see also Furtak et al. 2023).

Here we report the discovery of the first known AGN at $z > 8$. The galaxy harboring this AGN was first identified as a candidate $z \sim$ 8 galaxy by Roberts-Borsani et al. (2016a; named EGSY-2008532660). Its spectroscopic redshift was measured via Ly$\alpha$ emission via Keck/MOSFIRE to be $z_{Ly\alpha}$ = 8.683$^{+0.001}_{-0.004}$ from Zitrin et al. (2015; as EGSY8p7); at the time, and for several years, it was the farthest known Ly$\alpha$ emitter. It later gained a potential N V detection at $z_{sys}$ = 8.667, from Mainali et al. (2018), from additional Keck/MOSFIRE data, and it was also the brightest $z > 8.5$ galaxy found in the CANDELS survey (Finkelstein et al. 2022a; EGS_z910_6811). Here we report the results from JWST with the Cosmic Evolution Early Release Science (CEERS) Survey (Finkelstein et al. 2022b) data set. The NIRSpec spectroscopy (Arrabal Haro et al. 2023, in preparation) of this source has an ID from the micro shutter array (MSA) of 1019; thus, and hereafter, we refer to it as CEERS_1019.

In Section 2, we present the data from four different JWST observational modes from the CEERS Survey. In Section 3, we describe our method of emission-line fitting, while we explore the detected emission lines in detail in Section 4. In Section 5, we analyze the properties of this galaxy from the available imaging data and from our spectroscopic data in Section 6. We discuss the implications of these results in Section 7 and present our conclusions in Section 8. Throughout this paper we assume a flat Planck cosmology, with $H_0 = 67.36$ km s$^{-1}$ Mpc$^{-1}$, $\Omega_m = 0.3153$, and $\Omega_\Lambda = 0.6847$ (Planck Collaboration et al. 2020). All magnitudes are in the AB system, and all rest-frame wavelengths are vacuum.

## 2. Data

Data presented in this work were taken as part of the CEERS Survey (ERS 1345; PI: S. Finkelstein; Bagley et al. 2023; Finkelstein et al. 2022b) in the CANDELS (Grogin et al. 2011; Koekemoer et al. 2011) Extended Groth Strip (EGS) field. The complete details of the CEERS program will be presented in Finkelstein et al. (2023, in preparation), and the program data can be found at doi:10.17909/z7p0-8481 and at https://archive.stsci.edu/hlsp/ceers. This source is one of the first to be observed and published with four JWST (Gardner et al. 2023) observing modes: NIRSpec (Böker et al. 2023), NIRCam (Rieke et al. 2023), MIRI (Wright et al. 2023), and NIRCam/Wide-Field Slitless Spectrograph, or WFSS (Greene et al. 2017). We describe these observations below and provide a summary of information about this source in Table 1.

Additional imaging data for this source were obtained in the rest-frame infrared from Spitzer/MIPS at 24 μm and Herschel/PACS at 100 μm, as well as SCUBA-2 at 850 μm and the JVLA at 3 GHz, which are shown in Appendix A. X-ray imaging from the Chandra Space Observatory for this source is also discussed in Appendix B.

### 2.1. NIRSpec Observations

The source presented in this work is included in the JWST/NIRSpec (Jakobsen et al. 2022) multi-object shutter (MOS) configurations taken with the MSA (Ferruit et al. 2022) during the CEERS Epoch 2 observations (2022 December). These NIRSpec observations are split into six different MSA pointings, each of them observed with the G140M/F100LP, G235M/F170LP, and G395M/F290LP medium-resolution ($R \approx 1000$; here denoted by "M") gratings plus the prism ($R \approx 30$–300), fully covering the $\sim$1–5 μm wavelength range. The MSA was configured to use three-shutter slitlets, enabling a three-point nodding pattern, shifting the pointing by a shutter length plus the size of the bar between the shutters in each direction for background subtraction. The total exposure time per disperser is 3107 s, distributed as three integrations (one per nod) of 14 groups each in the NRSIRS2 readout mode.

JWST/NIRSpec 2D+1D spectra for this source in each of the M gratings (G140M, G235M, and G395M) are shown in

**Table 1**
Source Information for CEERS_1019

| | | | |
|---|---|---|---|
| R.A. | 215.0353914 | [deg] | (1) |
| Decl. | 52.8906618 | [deg] | (2) |
| $m_{F160W}$ | $25.2 \pm 0.03$ | [AB mag] | (3) |
| $m_{F356W}$ | $24.8 \pm 0.01$ | [AB mag] | (4) |
| $z_{phot}$ | $8.84^{+0.12}_{-0.25}$ | (HST+IRAC) | (5) |
| $z_{phot}$ | $8.72^{+0.04}_{-0.06}$ | (+JWST) | (6) |
| $z_{spec(Ly\alpha)}$ | $8.683^{+0.001}_{-0.004}$ | (Keck) | (7) |
| $z_{spec(Ly\alpha)}$ | $8.6854 \pm 0.0045$ | (NIRSpec) | (8) |
| $z_{spec([O\,III])}$ | $8.6788 \pm 0.0002$ | (NIRSpec) | (9) |
| $\log(M_\star)$ | $9.5 \pm 0.3$ | [$M_\odot$] | (10) |
| $\log(sSFR)$ | $-7.9 \pm 0.3$ | [$M_\odot$ yr$^{-1}$] | (11) |
| $\log(M_{BH})$ | $6.95 \pm 0.37$ | [$M_\odot$] | (12) |
| $T_e([O\,III])$ | $18,630.76 \pm 3.68$ | [K] | (13) |
| $12+\log(O/H)$ | $7.66 \pm 0.51$ | | (14) |
| Z | $0.095^{+0.21}_{-0.06}$ | [$Z_\odot$] | (15) |
| $n_e$ | $1.9 \pm 0.2 \times 10^3$ | [cm$^{-3}$] | (16) |
| UV Slope $\beta$ | $-1.76^{+0.12}_{-0.13}$ | | (17) |
| $A_v$ | $0.4 \pm 0.2$ | [mag] | (18) |

**Note.** Column (1): right ascension. Column (2): declination. Column (3): observed AB magnitude in the F160W filter from Finkelstein et al. (2022a). Column (4): observed AB magnitude in the F356W filter (Section 2.2). Column (5): photometric redshift measured with HST (prior to JWST photometry; Finkelstein et al. 2022a). Column (6): photometric redshift measured including JWST photometry from CEERS (Section 5). Column (7): spectroscopic redshift measured from Keck/MOSFIRE spectroscopy via Lyα emission-line detection from Zitrin et al. (2015). Column (8): spectroscopic redshift measured from JWST/NIRSpec spectroscopy via Lyα. Column (9): spectroscopic redshift measured from JWST/NIRSpec spectroscopy via [O III] emission-line detection (Section 4.1). Column (10): stellar mass from Prospector SED fit (Section 5). Column (11): specific star formation rate (Section 5). Column (12): BH mass (Section 6.1). Column (13): electron temperature (Section 6.2). Column (14): $T_e$-based metallicity (Section 6.2). Column (15): metallicity (Section 6). Column (16): electron density (Section 6.2). Column (17): UV spectral slope (Section 5.1). Column (18): dust attenuation (Sections 5 and 6.4).

Figure 1, with the location of the source marked by a horizontal red line in each 2D spectrum (see Appendix C for details of the prism observations that were corrupted). Details of the extraction method from 2D to 1D are detailed in Section 2.1.1 below.

### 2.1.1. NIRSpec Data Reduction

The details of the CEERS NIRSpec data processing are presented in Arrabal Haro et al. (2023). We summarize the main steps of the reduction here. The NIRSpec data are processed with the Space Telescope Science Institute Calibration Pipeline[42] (version 1.8.5; Bushouse et al. 2022a) and the Calibration Reference Data System (CRDS) context jwst_1027.pmap. We use the calwebb_detector1 pipeline module to subtract the bias and the dark current, correct the 1/f noise, and generate count-rate maps (CRMs) from the uncalibrated images. At this stage, the parameters of the jump step are modified for an improved correction of the "snowball" events[43] associated with high-energy cosmic rays.

---

[42] https://jwst-pipeline.readthedocs.io/en/latest/index.html
[43] https://jwst-docs.stsci.edu/data-artifacts-and-features/snowballs-and-shower-artifacts

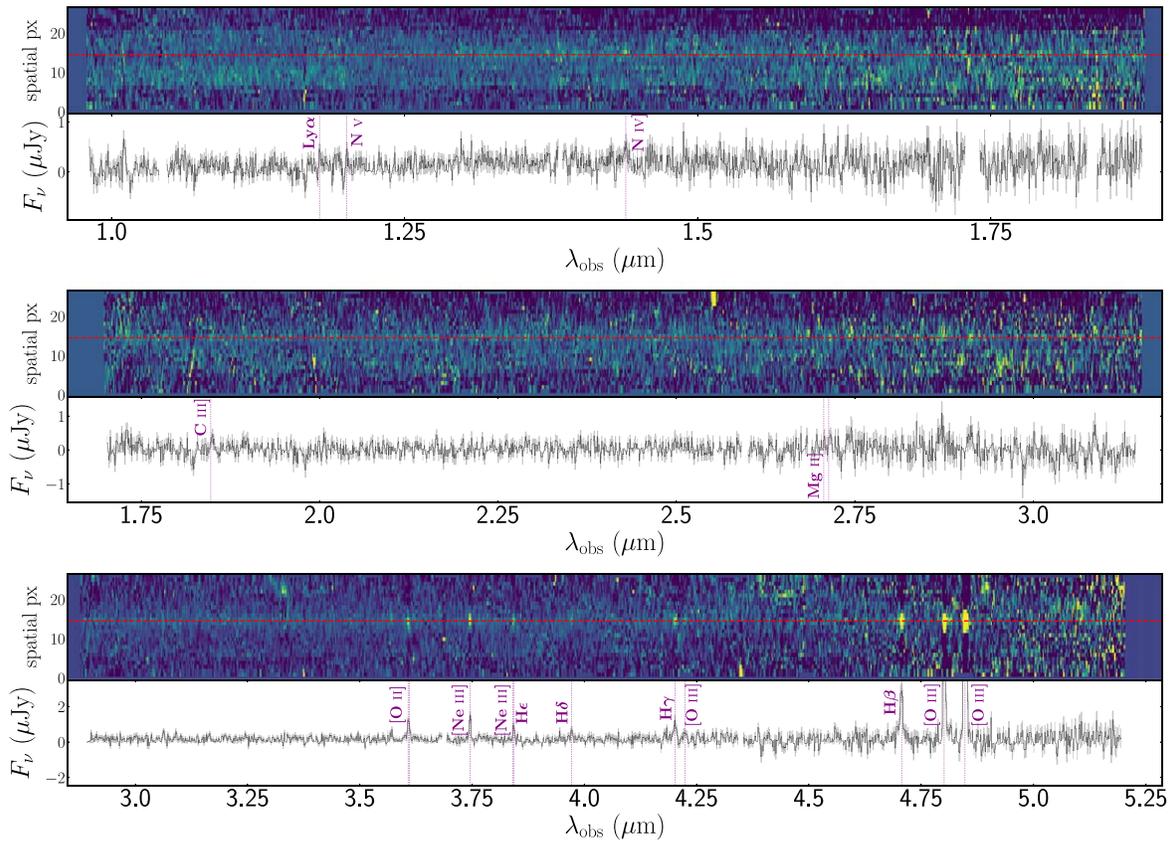

**Figure 1.** 2D and 1D spectra of CEERS_1019 from three JWST/NIRSpec M gratings: G140M (top), G235M (middle), and G395M (bottom). The horizontal red dashed line identifies the central location of the source in the 2D spectrum and is the extraction center for our 1D spectra. Measured emission lines are indicated with purple dotted lines and discussed in Section 4. A description of the CEERS NIRSpec observations for this source is given in Section 2.1, and the data reduction process is described in Section 2.1.1.

The resulting CRMs are then processed with the cal-webb_spec2 pipeline module, which creates two-dimensional (2D) cutouts of the slitlets, performs the background subtraction making use of the three-nod pattern, corrects the flat fields, implements the wavelength and photometric calibrations, and resamples the 2D spectra to correct the distortion of the spectral trace. The pathloss step accounting for the slit-loss correction is turned off at this stage of the reduction process (see Figure 2). Instead, we introduce slit-loss corrections based on the morphology of the sources in the NIRCam bands and the location of the slitlet hosting them.

The one-dimensional (1D) spectra of the sources were obtained via an optimal extraction (Horne 1986) with a spatial weight profile from the trace of the source, such that the pixels near the peak of the trace are maximally weighted. To create the extraction profile, the 2D signal-to-noise ratio (S/N) spectrum was collapsed in the spectral direction for each grating independently, taking the median value at each spectral pixel and fitting a Gaussian to the positive trace. This source is effectively unresolved at all wavelengths (the intrinsic size FWHM is less than the JWST point-spread function, or PSF, FWHM), so the central trace is unaffected by upper and lower negative traces. A significant trace was measured in both the G140M and G395M gratings, but was not as apparent in the G235M grating such that it could be fitted with a similar Gaussian profile. We thus used the profile from the G140M grating for the extraction from 2D to 1D in the G235M grating.

We evaluate the scaling factor for the slit-loss correction compared to the NIRCam photometry for this source, again using the reduction without the path-loss correction applied. In all NIRCam filters, we repeatedly calculate the fluxes enclosed within the $0.''2 \times 0.''46$ rectangle aperture, using the source and MSA shutter positions, and estimate the scaling factors to match them with the total flux measurements. The scaling factors range from approximately 2.0–2.5 among the NIRCam filters. We correct the spectrum with the scaling factor of the filter whose central wavelength is closest to that of the observed wavelength. This self-consistently applies an aperture correction accounting for the variable PSF across the observed wavelength range. We assess the uncertainty in these slit losses by comparing them to another method based solely on the F444W filter for multiple sources in the CEERS observations. We find differences in the correction factor on the order of 10%–40% (Fujimoto et al. 2023) on average for all sources. We note that the slit-loss correction is multiplicative, such that any systematic uncertainty in this correction does not affect the significance measurements for emission lines.

We test the accuracy of the error spectrum computed by the pipeline by comparing the normalized median absolute deviation of the science spectrum (masking out emission lines and removing a smoothed continuum) to the median of the error spectrum in each grating individually. The real data show fluctuations ∼1.5–2 times larger than the typical error value. Thus, we measure and scale the error spectrum up by this scale factor in each grating.

Once the errors are corrected, and the spectra are scaled to the NIRCam photometry, the three M gratings are combined into a single spectrum, resampling to a common wavelength array at the overlapping wavelengths and adopting the mean flux at each pixel, weighted by the flux errors. This combined spectrum is then used for the remainder of the analysis in this paper.

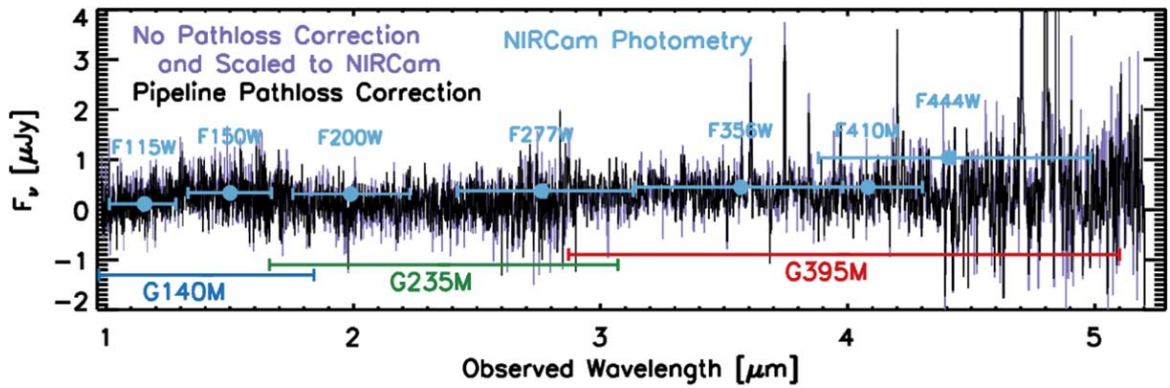

**Figure 2.** Comparison of the JWST/NIRSpec combined M grating spectrum when using the native pipeline `pathloss` correction (black) vs. without this step of the pipeline, but rather scaling the spectra to the measured JWST/NIRCam photometry (blue), as described in Fujimoto et al. (2023) and in Section 2.1.1. Our scaled spectrum (purple) highlights how the default pipeline `pathloss` correction underpredicts the total slit-loss correction for this source and that corrections to the NIRSpec spectrum are required for the flux calibration of resolved sources.

### 2.2. NIRCam Imaging

The galaxy discussed in this work was observed in the CEERS JWST/NIRCam (Rieke et al. 2003, 2005; Beichman et al. 2012; Rieke et al. 2023) imaging taken in 2022 December (in CEERS NIRCam Field 8); these imaging data, including the detailed reduction steps, are described in Bagley et al. (2023).[44] Photometry of this source was measured using Source Extractor (SE; Bertin & Arnouts 1996) in dual-image mode. The photometry procedure is broadly similar to that described in full by Finkelstein et al. (2022b), though with a few differences we detail here. First, to improve color accuracy, especially in the Hubble Space Telescope (HST)/WFC3 bands, we do a two-step PSF correction. For any filters with a PSF FWHM smaller than the NIRCam F277W (which includes HST/Hubble's Advanced Camera for Surveys (ACS), F606W and F814W and NIRCam F115W, F150W, and F200W), we convolve the images with a kernel designed to match these images' PSFs to that in F277W. For images with larger PSFs (WFC3 F105W, F125W, F140W, and F160W and NIRCam F356W, F410M, and F444W), we derive a correction factor. This is done by convolving the F277W image with a kernel designed to match the PSF to the larger PSF image, with a correction factor derived as the ratio of the flux in the broadened image to that in the original F277W image. In this way, we correct for light not captured in our default aperture in the bands with larger PSFs, without needing to smooth all bands to that with the largest PSF (which would be WFC3 F160W).

We tested this process using source injection simulations to confirm that accurate colors were recovered. Finally, following Finkelstein et al. (2022b), we derived accurate total fluxes, first by correcting the fluxes in all bands by an aperture correction derived in F277W (as the ratio between the flux in our small Kron aperture used to measure colors and in the default larger "FLUX_AUTO" Kron aperture), and then correcting this by a factor of 8%–10%, to account for missing light from the wings of the PSF, with this correction factor derived via the source injection simulations (see Finkelstein et al. 2022b for more details).

This final photometry catalog includes measurements over the full CEERS NIRCam wavelength range in the F115W, F150W, F200W, F277W, F356W, F410M, and F444W filters, which have exposures of ∼3000 s per filter (∼6000 s for F115W), as well as in the existing HST/CANDELS ACS and WFC3 F606W, F814W, F105W, F125W, F140W, and F160W bands. We include the F098M (nondetection) data for this source from Finkelstein et al. (2022a).

The measured photometric redshift for this galaxy before JWST was $z_{\mathrm{phot}}(HST) = 8.84^{+0.12}_{-0.25}$ (Finkelstein et al. 2022a). With the addition of the JWST/NIRCam imaging (and by removing the blended Spitzer/IRAC data), we measure a photometric redshift with EAZY (using the same process as Finkelstein et al. 2022b) nearly equal to the spectroscopic redshift ($z_{\mathrm{spec}} = 8.679$) of $z_{\mathrm{phot}}(HST + JWST) = 8.68^{+0.09}_{-0.15}$.

### 2.3. MIRI Imaging

This source was also observed with JWST/MIRI in CEERS Epoch 1 with filters F560W and F770W. The photometry is obtained from the CEERS team DR0.5 data release for the MIRI 3 and MIRI 6 fields (G. Yang et al. 2023, in preparation).[45] The MIRI data reduction process is described by Yang et al. (2021),[46] and the photometric measurements for this source have previously been reported by Papovich et al. (2023). We present the JWST/NIRCam and MIRI photometry for this source in Table 2. With the addition of the MIRI photometry, the photometric redshift is slightly modified to $z_{\mathrm{phot}}(HST + JWST) = 8.72^{+0.04}_{-0.06}$.

### 2.4. NIRCam Grism Observations

This source was also observed with the JWST/NIRCam WFSS in December 2022 as part of the CEERS Epoch 2 data. These data were obtained using the F356W filter with the orthogonal column "C" and row "R" grisms with a resolution $R \sim 1600$. Grism spectra for this source are shown in Figure 3;

---

[44] Specifically, JWST/NIRCam imaging data were reduced using the JWST pipeline, Version 1.8.5 (Bushouse et al. 2022a), and CRDS context `jwst_1023.pmap`. See also the CEERS team DR0.6 data release.

[45] https://ceers.github.io/releases.html
[46] Specifically, JWST/MIRI imaging data were reduced using the JWST pipeline, Version 1.7.2 (Bushouse et al. 2022b), and CRDS context `jwst_1027.pmap`.

**Table 2**
Photometric Measurements for CEERS_1019, with Fluxes in nJy

| Instrument | Filter | Flux (nJy) |
|---|---|---|
| HST/ACS | F606W | $9.0 \pm 7.7$ |
|  | F814W | $1.0 \pm 8.1$ |
| HST/WFC3 | F098M | $-11.0 \pm 18.4$ |
|  | F125W | $200.8 \pm 11.3$ |
|  | F140W | $272.1 \pm 19.2$ |
|  | F160W | $306.2 \pm 9.7$ |
| JWST/NIRCam | F115W | $117.1 \pm 3.8$ |
|  | F150W | $334.2 \pm 7.1$ |
|  | F200W | $313.7 \pm 5.6$ |
|  | F277W | $371.4 \pm 4.5$ |
|  | F356W | $449.6 \pm 3.5$ |
|  | F410M | $450.1 \pm 8.8$ |
|  | F444W | $1039.2 \pm 7.7$ |
| JWST/MIRI | F560W | $426.8 \pm 21.6$ |
|  | F770W | $404.0 \pm 16.9$ |

**Note.** The HST photometry is updated from Finkelstein et al. (2022a) using the NIRCam-selected apertures as described in Finkelstein et al. (2022b). JWST/NIRCam photometry from CEERS Epoch 2 imaging measured in a similar way as Finkelstein et al. (2022b), described in Section 2.2. JWST/MIRI photometry from Papovich et al. (2022) using CEERS Epoch 1 imaging is described in Section 2.3.

the data reduction methodology will be presented in N. Pirzkal et al. (2023, in preparation).[47] We observe several emission lines detected in the combined grism spectrum. These are consistent with [O II], [Ne III], H$\epsilon$, and H$\delta$ if a wavelength shift (of ∼40 Å) is applied from the NIRCam to NIRSpec spectra. This is likely due to as yet improving wavelength calibrations for both instruments, though given the agreement of our observed Ly$\alpha$ wavelength with ground-based measurements, the offset is likely dominated by the NIRCam WFSS spectra. While upcoming improved calibrations will provide the necessary adjustments to compare better the resulting line fluxes and detections from JWST, these detected lines show the utility of this NIRCam WFSS mode for early galaxy spectroscopy. For the remainder of this paper, we use all measurements from the higher-S/N NIRSpec spectra.

### 3. Methods: Emission-line Search

To search for emission-line features in our 1D spectrum, we utilize an automated line-finding code first published in Larson et al. (2018) and outlined here. This code uses a Monte Carlo Markov Chain (MCMC) routine to fit a model that consists of a Gaussian line plus a continuum constant to a given wavelength range, with four free parameters: the continuum level, line central wavelength, line FWHM, and integrated line flux. To run the MCMC, we use an IDL implementation of the affine-invariant sampler (Goodman & Weare 2010) to sample the *posterior* similar to that used in Finkelstein et al. (2019), which is similar to the Python `emcee` package (Foreman-Mackey et al. 2013).

As a first pass-through for emission-line features, we search the entire spectrum with our automated code. At each wavelength pixel, we do an initial S/N check (flux/error). If this is greater than unity, we run our fitting routine (this "precheck" is not required, but helps with efficiency, as no detectable emission feature would have S/N < 1 at the line center). We use a fitting range of 100 times the pixel scale in the G140M grating on either side of the center pixel, which equates to 630 Å across the full spectrum. This ensures we fit the same wavelength range for every line, regardless of pixel scale. We fit a single Gaussian feature to the spectrum:

$$f(\lambda) = f_c + f_0 \exp\left(-\frac{1}{2}\frac{(\lambda - \lambda_0)^2}{\sigma^2}\right). \quad (1)$$

We also impose noninformative priors on each free parameter, designed to increase the efficiency of the line detections while removing the chance for the MCMC chain to exit a range of realistic parameters. For the continuum constant ($f_c$), we let it vary between $-3$ to 10 times the average flux across the fitting range. These values are larger than the $1\sigma$ noise level and broadly encompass the typical continuum values for our source. We restrict the peak wavelength ($\lambda_0$) to be the wavelength at that pixel $\pm$ one pixel, such that we fit a Gaussian within each pixel. The actual value in Å varies, as the pixel scale ($\Delta\lambda$) for NIRSpec is wavelength-dependent, with $\Delta\lambda_{G140M} = 6.4$ Å, $\Delta\lambda_{G235M} = 10.7$ Å, and $\Delta\lambda_{G395M} = 17.9$ Å. We limit the FWHM (FWHM $=2\sqrt{2\ln(2)}\,\sigma = 2.355\sigma$ to $\pm$ 30 km s$^{-1}$ of the observed FWHM of the [O III]$_{5008}$ line, the brightest feature in the spectrum (see Figure 4 and Table 3). Tying the FWHM to [O III]$_{5008}$ in velocity units rather than pixels is critical, as the wavelength dispersion changes across the observed wavelength range. The line flux prior requires the line flux ($f_0$) to be greater than the average of the flux over the plotting range after a $>1\sigma$ clipping (estimate of the continuum at that location) and less than 100 times the average flux, well outside the expected range of line fluxes.

To define the starting point for our MCMC routine, we used the IDL routine `mpfit`, a Levenberg–Marquardt least-squares fitting routine that fits a Gaussian with the above parameters (Markwardt 2009). We then run our MCMC fitting code with 10,000 iterations and 100 walkers on each pixel and determine the best line fit results as outlined in the following sections. We use the median of the last 10,000 steps of our MCMC chain for our fit parameters. To measure the MCMC error on our parameters, we use the `robust sigma` calculation: using the median absolute deviation as the initial estimate, then weighting points using Tukey's biweight (Equation (9) from Beers et al. 1990), assuming a Gaussian posterior distribution. As our MCMC fits marginalize over the line flux and the continuum flux, the errors on both are included in our final line fit error.

We place several constraints on the resulting Gaussian fits when determining a successful line detection, including the wavelength and comparison to neighboring pixels. First, we mask out the edges of the wavelength range for the three M Gratings (within 0.97–5.1 $\mu$m), where the grating transmission curve falls below 70%. To determine the correct peak pixel for the line, we select the fit at the pixel that has the highest peak flux (the flux at the peak of the Gaussian fit) within the FWHM of the line. We calculate an integrated S/N from the MCMC Gaussian fit as the median line flux divided by the robust sigma of the line flux as described above. We also measure a peak S/N value, calculated by taking the maximum flux of the Gaussian fit and calculating the ratio of this to the noise per pixel within our fitting range (similar to Larson et al. 2022). For our initial search through the spectrum for emission lines, we only consider lines found at >5$\sigma$ in both integrated and peak S/N. This initial pass is run for masking out significant features in the spectrum to fit and remove the detected continuum, as described in Section 3.1, and for identifying potentially unexpected emission-line detections in sources at this redshift. In Section 3.2, we discuss how these initial S/N measurements overestimated the significance of our emission-line detections and how we determine the final line flux errors and S/N values for our detected emission lines.

---

[47] Briefly, we used CRDS `jwst_1089.pmap` up to stage 1 of the reduction (flat fields and world coordinate system, WCS); all further processing was our own, and the WFSS configuration was custom from 2023 May.

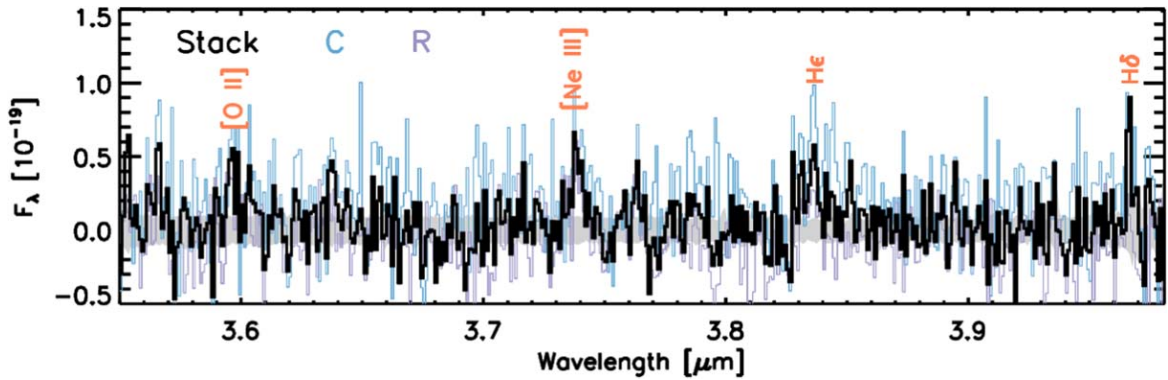

**Figure 3.** The spectrum of CEERS_1019 obtained with JWST/NIRCam's wide-field slitless spectroscopy mode. This consists of spectra with both the column ("C"; blue) and row ("R"; purple) grisms, both taken with the F356W blocking filter, with a combined spectrum shown in black. While these data have a higher noise level than the NIRSpec spectra, we observe detections of the same [O II]$_{3727+3729}$, [Ne III]$_{3870}$, and bluer Balmer lines that we see in NIRSpec (Figure 5). While we do not use the grism measurements in this paper, these data highlight the utility of this mode for obtaining slitless measurements of modestly faint emission lines out to high redshifts.

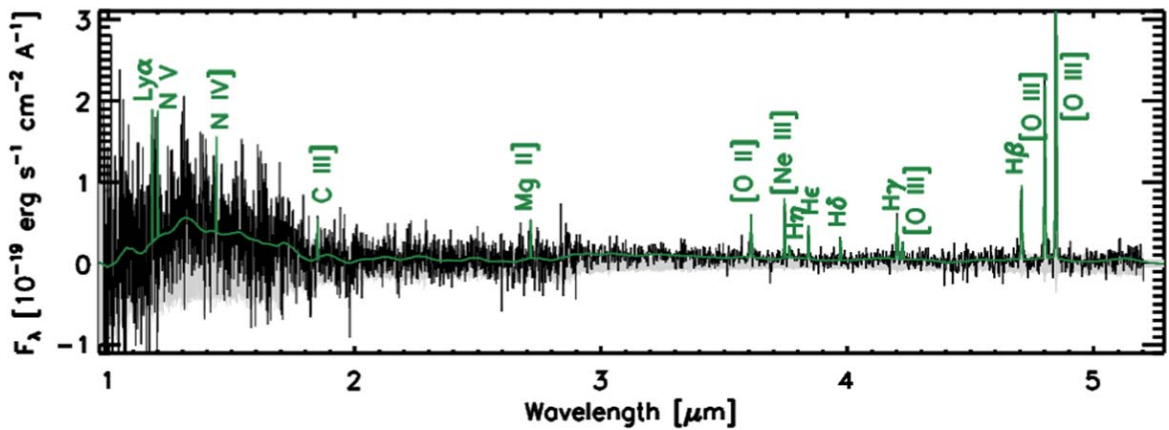

**Figure 4.** Combined JWST/NIRSpec spectrum from G140M+G235M+G395M of CEERS_1019, plotted in $F_\lambda$ [$10^{-19}$ erg s$^{-1}$ cm$^{-2}$ Å$^{-1}$] vs. observed wavelength [$\mu$m]. Plotted in green is the fit to the continuum in the spectrum (see Section 3.1) plus the detected emission lines, as discussed in Section 4. We require an S/N >2.3 (with our simulation-estimated noise as described in Section 3.2) for a line to be considered detected.

### 3.1. Measuring Continuum in NIRSpec

To fit the continuum to the NIRSpec spectra, we first mask out all the emission lines (the peak of the line ± FHWM) and interpolate over those pixels with an average of the three not-masked pixels on either side of the line region. We then smooth the entire spectrum using a boxcar filter with a width of 60 pixels. As each M grating has a different pixel scale, we then smooth each section of the spectrum with a larger boxcar filter at the blue end compared to the red end. We use 120 pixels where $\lambda < 1.75\,\mu$m (the G140M grating), 60 pixels from $1.75 < \lambda < 2.9\,\mu$m (the G235M grating), and 30 pixels where $\lambda > 2.9\,\mu$m (the G395 grating). To remove any discontinuous jumps between the gratings, we smooth the whole spectrum again with the 60 pixel boxcar filter. This measured continuum is plotted in green over the combined NIRSpec spectrum in Figure 4, with notable detection in the G140M grating but nonsignificant detection in the redder gratings. This estimated continuum is consistent with the observed photometric spectral energy distribution (SED), as expected due to our slit-loss correction methodology described in Section 2.1.1.

### 3.2. Determining the Significance of the Emission Lines

Upon inspection of the initially identified lines described above, it was found that this methodology tended to overestimate the S/N, especially for faint lines. We thus empirically derive line flux errors and appropriate S/N measurements for each line that passes the above selection cuts in the following way. Using the line-masked spectrum as above, we find the 20 closest "blank" pixels, free of any other nearby detected emission lines, on either side of the emission line (line center + FWHM, such that we are not overlapping the emission line at all). At each of these blank pixels, we insert a fake emission line with the same parameters as our detected emission line ($F(\lambda)$, FHWM, and $f_c$) and a line center ($\lambda_0$) at the wavelength of that pixel. We then run the same line-fitting routine at each pixel and record the recovered line flux from the MCMC ($F(\lambda)_{\rm out}$) for each of our 40 simulated lines. Our reported line flux error ($\Delta F(\lambda)$) is then the median absolute deviation of these recovered line fluxes, as reported

**Table 3**
Measured Line Fluxes in CEERS_1019

| Line $\lambda_{rest}$ (1) | $\lambda_{obs}$ Central/Blue [Å] (2) | S/N (3) | Line Flux Total [$10^{-18}$ cgs] (4) | FWHM Single/Blue [km s$^{-1}$] (5) | Flux Narrow/Blue [$10^{-18}$ cgs] (6) | Flux Broad/Red [$10^{-18}$ cgs] (7) | FWHM Broad/Red [km s$^{-1}$] (8) | Rest EW [Å] (9) | $\Delta v$ [km s$^{-1}$] (10) |
|---|---|---|---|---|---|---|---|---|---|
| Ly$\alpha_{1215}$ | 11775.10 ± 6.58 | 14.4 | 5.41 ± 0.38 | 209.77 ± 121.21 | ⋯ | ⋯ | 1196.74 ± 470.98 | 10.67 ± 0.93 | 218.61 ± 332.94 |
| N V$_{1238+1242}$ | 11969.40 ± 3.39 | 2.0 | 1.74 ± 0.88 | 358.93 ± 19.66 | 0.03 ± 0.03 | 1.70 ± 0.72 | ⋯ | 3.90 ± 2.30 | −522.45 ± 85.20 |
| N IV]$_{1483+1486}$ | 14357.60 ± 8.77 | 7.6 | 3.36 ± 0.44 | 363.99 ± 20.77 | 1.23 ± 0.85 | 2.13 ± 0.69 | ⋯ | 7.96 ± 1.47 | 16.15 ± 183.26 |
| [O II]$_{3727+3729}$ | 36074.80 ± 4.09 | 11.9 | 2.94 ± 0.25 | 357.55 ± 20.83 | 1.80 ± 0.35 | 1.15 ± 0.35 | ⋯ | 34.79 ± 5.80 | 11.43 ± 34.71 |
| [Ne III]$_{3869}$ | 37456.30 ± 2.29 | 12.84 | 3.5 ± 0.27 | 358.44 ± 17.43 | ⋯ | ⋯ | ⋯ | 44.27 ± 7.07 | 8.51 ± 19.64 |
| He I$_{3889}$+H$\eta_{3890}$ | 37656.80 ± 9.73 | 2.3 | 0.66 ± 0.29 | 357.13 ± 18.12 | ⋯ | ⋯ | ⋯ | 8.43 ± 3.87 | 71.95 ± 77.78 |
| [Ne III]$_{3968}$+H$\epsilon_{3971}$ | 38407.90 ± 3.75 | 5.5 | 2.04 ± 0.37 | 347.65 ± 21.01 | 0.48 ± 0.43 | 1.56 ± 0.40 | ⋯ | 26.53 ± 6.07 | −23.34 ± 30.11 |
| H$\delta_{4102}$ | 39717.90 ± 6.07 | 3.1 | 1.14 ± 0.37 | 358.56 ± 18.29 | ⋯ | ⋯ | ⋯ | 15.64 ± 5.50 | 53.32 ± 46.36 |
| H$\gamma_{4341}$ | 42018.20 ± 5.25 | 6.6 | 2.60 ± 0.39 | 360.56 ± 18.78 | ⋯ | ⋯ | ⋯ | 39.36 ± 8.48 | −27.63 ± 38.12 |
| [O III]$_{4364}$ | 42244.90 ± 9.91 | 3.0 | 1.16 ± 0.38 | 361.11 ± 17.79 | ⋯ | ⋯ | ⋯ | 17.78 ± 6.45 | 19.06 ± 70.68 |
| H$\beta_{4862}$ | 47059.80 ± 4.91 | 10.2 | 7.32 ± 0.72 | 354.80 ± 19.61 | 3.79 ± 0.69 | 3.53 ± 1.51 | 1196.26 ± 349.07 | 128.94 ± 18.96 | −31.55 ± 32.06 |
| [O III]$_{4960}$ | 48009.60 ± 1.87 | 26.3 | 13.08 ± 0.50 | 358.52 ± 17.37 | ⋯ | ⋯ | ⋯ | 235.35 ± 27.51 | 1.00 ± 13.64 |
| [O III]$_{5008}$ | 48473.50 ± 1.14 | 47.2 | 39.58 ± 0.84 | 358.66 ± 12.23 | ⋯ | ⋯ | ⋯ | 725.59 ± 82.2 | −0.59 ± 124.41 |
| C III]$^{*}_{1908}$ | 18493.30 ± 9.22 | 1.63 | 1.70 ± 1.05 | 642.90 ± 156.84 | ⋯ | ⋯ | ⋯ | 6.04 ± 3.85 | 309.51 ± 149.63 |
| Mg II$_{2796+2803}$ | 27077.90 ± 6.67 | 1.18 | 1.62 ± 1.37 | 358.76 ± 20.82 | 0.67 ± 1.01 | 0.96 ± 1.20 | ⋯ | 11.56 ± 9.90 | 144.74 ± 74.18 |

**Note.** Identified emission-line values in CEERS_1019; see Figure 5 for individual plots. Column (1): line name and rest-frame vacuum wavelength in [Å]. Column (2): observed wavelength of the line in NIRSpec in [Å]. Column (3): S/N of the line (Section 3.2). Column (4): total integrated line flux in $F_\lambda$ (units of $10^{-18}$ erg s$^{-1}$ cm$^{-2}$ Å$^{-1}$; Section 3). Column (5): FWHM of the line fit (in km s$^{-1}$) for single-Gaussian fits, the blue line in the doublet (Section 3.4.2) and asymmetric (Section 3.4.1) fits, or the narrow component in the dual-component (Section 3.4.3) fits. Column (6): flux of the line fit (in $10^{-18}$ erg s$^{-1}$ cm$^{-2}$ Å$^{-1}$) for single-Gaussian fits, the blue line in the doublet (Section 3.4.2) and asymmetric (Section 3.4.1) fits, or the narrow component in the dual-component (Section 3.4.3) fits. Column (7): flux of the red line in the doublet (Section 3.4.2) fit or the broad component in the dual-component (Section 3.4.3) fit (in $10^{-18}$ erg s$^{-1}$ cm$^{-2}$ Å$^{-1}$). Column (8): FWHM of the red line in the asymmetric (Section 3.4.1) fit or the broad component in the dual-component (Section 3.4.3) fit (in km s$^{-1}$). Column (9): rest-frame EW (Rest EW) of the line (in Å; Section 3.3). Column (10): velocity offset of the emission line from the systemic redshift of CEERS_1019 ($z_{[O III]}$=8.679; Section 4.1; in km s$^{-1}$). Lines detected at a low significance are reported below the horizontal line and are included for completeness. $^{*}$While the C III] line is a doublet with peaks at 1906 and 1908 Å in the rest frame, we do not resolve this doublet and show the best fit, which uses a single-Gaussian profile with a broad FHWM (Section 4).

in column 4 of Table 3. The S/N of our emission lines is taken as the measured line flux from our line fit divided by this error, S/N = $F(\lambda)/\Delta f(\lambda)$, and is reported in column 3 of Table 3. To derive a consistent threshold in S/N where we consider a line to be detected, we examine the results of the automated line-finding routine, searching for lines with rest-frame wavelengths that do not match a known line (using a large line list). The point below which we detect no "unknown" (spurious) emission lines is S/N = 2.3, which we use as our detection threshold for emission lines in CEERS_1019, unless otherwise noted below.

### 3.3. Rest-frame Equivalent Width Measurements

To measure rest-frame equivalent widths (EWs) for our emission lines (EW$_{rest} = \frac{F(\lambda)}{f_c(1+z)}$), we require a measurement of the continuum emission in the region near a given emission line. While, in principle, this could come from the spectrum itself, the continuum is only detected at a high significance in the G140M (blue) grating. Thus, to enable a uniform procedure for all emission lines, we use the best-fit Prospector model (see Section 5 and Figure 7) for our rest-frame EW measurements. We note that this is based on the same NIRCam photometry used to derive our slit-loss corrections, and we also find that the agreement with the spectroscopic continuum in G140M is excellent. To determine the continuum value of the Prospector model, we first mask out any emission lines and smooth the spectrum following the same process described above. Our EW continuum value is the median of the Prospector continuum over the width of the line. The error is taken as the median absolute deviation of the Prospector model over this same range.

### 3.4. Non-Gaussian Emission-line Fits

In our automated line search above, we fit a single Gaussian to each emission line, which is not physical for many of the lines in our spectrum. We thus implement three additional line profiles: a dual-component Gaussian, a doublet Gaussian, and an asymmetric Gaussian fit as described below. The fits are done using the same MCMC routine as above, and any parameters not mentioned below as having been changed or added are the same as those in the single-Gaussian fits.

#### 3.4.1. Asymmetric Gaussian Fit

Given the detection of Ly$\alpha$ in our NIRSpec spectra, we want to fit an appropriate asymmetric Gaussian profile to the data to account for intergalactic medium (IGM) absorption on the blue side of the line. For this fit, we add an extra FWHM parameter, such that we now have "blue" and "red" FWHMs of the line. The blue FWHM can vary between 0, as the blue side of the line may be fully attenuated, and FWHM$_{[O\ III]}$ + 30 km s$^{-1}$, which is the same maximum as the single-Gaussian fit. We start the MCMC chain for this parameter at FWHM$_{[O\ III]}$. The red FWHM is less restricted, as it can vary between FWHM$_{[O\ III]}$ and 2000 km s$^{-1}$, and the MCMC chain is started at FWHM$_{[O\ III]}$ + 30 km s$^{-1}$, the maximum value for the blue FWHM. Following the same method as Jung et al. (2020), the equation for the asymmetric Gaussian is:

$$f(\lambda)_{\text{Asym}} = f_c + f_0 \times \begin{cases} \exp\left(-\frac{1}{2}\frac{(\lambda-\lambda_0)^2}{\sigma_{\text{blue}}^2}\right) & \text{for } \lambda \leqslant \lambda_0, \\ \exp\left(-\frac{1}{2}\frac{(\lambda-\lambda_0)^2}{\sigma_{\text{red}}^2}\right) & \text{for } \lambda > \lambda_0 \end{cases}. \quad (2)$$

This returns the continuum flux ($f_c$), peak line flux value ($f_0$), a peak wavelength ($\lambda_0$), and the blue- and red-side line widths ($\sigma_{\text{blue}}$ and $\sigma_{\text{red}}$). To measure our line flux, we integrate this emission-line profile and report the integrated line flux in column 4 of Table 3

#### 3.4.2. Doublet Gaussian Fits

Several of the UV and optical emission lines are not a single emission-line feature, but rather a doublet. To accurately fit these lines, we implement a doublet Gaussian line profile to our MCMC routine:

$$f(\lambda)_{\text{Doublet}} = f_c + f_1 \exp\left(-\frac{1}{2}\frac{(\lambda-\lambda_1)^2}{\sigma^2}\right) + f_2 \exp\left(-\frac{1}{2}\frac{(\lambda-(\lambda_1+\delta\lambda))^2}{\sigma^2}\right). \quad (3)$$

For this fit, we employ the continuum flux ($f_c$) plus two Gaussians with the same FWHM ($\sigma$) and individual line fluxes for each line ($f_1$ and $f_2$, respectively), with the blue peak ($\lambda_1$) fixed to ±6 pixels from the expected location and the red peak fixed to the redshifted separation between the two doublet lines ($\delta\lambda = \delta\lambda_{\text{rest}}(1+z)$). This returns a total combined line flux for the doublet ($f(\lambda)$) as reported in column 4 in Table 3, as well as individual line fluxes for each component, as reported in columns 6 and 7 of that same table.

#### 3.4.3. Dual-component Gaussian Fit

Upon inspection of the initial emission lines, the H$\beta$ line had a noticeable second broad component feature, as discussed in Section 4.2 below. In order to accurately measure this, we employ a dual-component Gaussian profile:

$$f(\lambda)_{\text{Dual}} = f_c + f_{\text{nar}} \exp\left(-\frac{1}{2}\frac{(\lambda-\lambda_0)^2}{\sigma_{\text{narrow}}^2}\right) + f_{\text{broad}} \exp\left(-\frac{1}{2}\frac{(\lambda-\lambda_0)^2}{\sigma_{\text{broad}}^2}\right). \quad (4)$$

This fit utilizes the continuum flux ($f_c$) plus two Gaussians with the same line center ($\lambda_0$), individual line fluxes for each line component ($f_{\text{nar}}$ and $f_{\text{broad}}$, respectively), and separate FWHMs for each ($\sigma_{\text{nar}}$ and $\sigma_{\text{broad}}$). The narrow FWHM has the same constraints as the single-Gaussian fit (±FHWM$_{[O\ III]}$), but the broad component is not allowed to be smaller than the narrow component (>FHWM$_{[O\ III]}$ + 30 km s$^{-1}$) and can extend to the unrestricted FWHM (<2000 km s$^{-1}$). We start the MCMC chain for the narrow FWHM at FHWM$_{[O\ III]}$ and the FWHM of the broad component at three times this value. This returns a total combined line flux for the line ($F(\lambda)$) as reported in column 4 in Table 3, as well as individual line fluxes for each component, as reported in columns 6 and 7 of that same table.

## 4. Emission-line Measurements

We find abundant emission lines in CEERS_1019 and present our measured line fluxes in Table 3. A plot indicating the detected emission lines is shown in Figure 4, and fits to each line are shown in Figure 5. In the following sections, we briefly discuss the fits to different lines and provide a detailed description (where necessary) of the detected emission lines. We note that H$\beta$ has a significant broad component, which we discuss in detail in Section 4.2.

### 4.1. Redshift Confirmation

We determine the spectroscopic redshift using the fit to the [O III]$_{5008}$ emission line, the brightest in our spectrum. This line is detected with a peak wavelength of $\lambda = 48473.50 \pm 1.14$ Å, yielding a spectroscopic redshift for this source of $z_{\text{[O III]}} = 8.6788 \pm 0.0002$. When comparing this to the redshift we measure from the line centers of the other strong [O III]$_{4960}$ line, we get a consistent measurement within $\Delta z/z \leqslant 10^{-4}$. We use this value as the systemic redshift for CEERS_1019.

### 4.2. Broad H$\beta$ Feature

The observed profile of the H$\beta$ emission line shows a strong narrow component, similar to [O III], but also a weaker broad component. To explore the significance of this broad component, we perform a dual Gaussian fit to this line, as described in Section 3.4.3. Uncertainties on both components were derived via similar simulations, as described in Section 3.2. From this fit, we find a distinct broad component to the H$\beta$ emission line, with an FWHM = $1196 \pm 349$ km s$^{-1}$. We show this fit (green) in Figure 6. The flux in this broad component is comparable to the flux in the narrow component, with a broad-to-narrow flux ratio of 0.93.

We consider multiple measures of the significance of this broad feature. The first is the S/N of the broadline flux from this fit, which is 2.5. However, this does not consider how good the fit is *without* the broadline component. We next try a narrowline-only fit, with the FWHM allowed to be free (i.e., not tied to [O III]), shown as the blue line in Figure 6. We find a single-line FWHM of $500 \pm 52$ km s$^{-1}$, significantly broader than the fiducial narrowline-only FWHM of $367 \pm 17$ km s$^{-1}$ (the purple line in Figure 6), making it clear that the broad component affects even a single-line fit.

Finally, we calculate the Bayesian Information Criterion (BIC) between the dual-component and narrowline-only fits. The BIC is a method of comparing the goodness-of-fit between two models, accounting for differing degrees of freedom. It is defined as BIC=$\chi^2 + k\,ln(N)$, where $k$ is the number of free parameters and $N$ is the number of data points (the number of spectral pixels, in this case; Liddle 2004). Our narrowline-only fit with four free parameters ($f_c$, $f_0$, $\lambda_0$, and $\sigma$—see Section 3) has a BIC of 21.4 when the FWHM is tied to [O III] and 21.8 when the FWHM is free. The narrow+broad fit with six free parameters ($f_c$, $\lambda_0$, $f_{\text{nar}}$, $f_{\text{broad}}$, $\sigma_{\text{nar}}$, and $\sigma_{\text{broad}}$—see Section 3.4.3) has a BIC of 18.7. The $\Delta$BIC=2.7–3.1 gives "positive" evidence (using the scale of Jeffreys 1961, where $\Delta$BIC > 6 is "strong" evidence) that a broadline component is necessary to fit the H$\beta$ line successfully.

This significant broad component of H$\beta$ thus indicates the presence of high-velocity gas, which we interpret as emitting from the broadline region of an AGN. Large-scale outflows can similarly produce broad velocity widths, but would be apparent in all emission lines (e.g., Amorín et al. 2012; Hogarth et al. 2020). We do not observe any broad features in lines with much higher S/N, like [O III]$\lambda$5008. Instead, [O III] and other forbidden lines have widths that are inconsistent with the outflow scenario and are instead consistent with the narrow H$\beta$ component. Broadline AGNs typically exhibit broad components for permitted lines (like H$\beta$) and narrow widths of forbidden lines (like [O III]; e.g., Schmidt et al. 2016; Vanden Berk et al. 2001), as observed in our spectrum. We discuss further evidence for the AGN nature of this source in Section 4.4 below.

### 4.3. Ly$\alpha$

The Ly$\alpha$ line was first detected by Zitrin et al. (2015) at $z_{Ly\alpha} = 8.683$ with Keck/MOSFIRE. We detect an emission line at this same wavelength, which we fit with an asymmetric Gaussian profile, as described in Section 3.4.1, and measure a line flux $f_{Ly\alpha} = 5.41 \pm 0.38 \times 10^{-18}$ erg s$^{-1}$ cm$^{-2}$. Using our measured central wavelength and the vacuum wavelength for Ly$\alpha$ of 1215.67 Å, we measure a Ly$\alpha$-based redshift of $8.6854 \pm 0.0045$, $218 \pm 333$ km s$^{-1}$ higher than that of [O III]. This is lower, although consistent within $\sim 1\sigma$, of the measured Ly$\alpha$ velocity offset (relative to N V) of $+362$ km s$^{-1}$ from Mainali et al. (2018).

We find a clear asymmetric line profile with an extended redside tail (FWHM$_{\text{red}} > 1000$ km s$^{-1}$). Although it becomes more difficult to detect Ly$\alpha$ at this high redshift, due to the increasingly neutral IGM, one can expect the escape of Ly$\alpha$ that is scattered and redshifted enough to avoid resonant scattering (Dijkstra et al. 2004), as seen in the extended red tail in our Ly$\alpha$ spectra. Additionally, a sharp cutoff at the blue edge of the line profile may indicate a significant contribution of resonant scattering, which can be caused by a proximate optically thick medium (Mason & Gronke 2020). Particularly, reionization simulations predict the sharp blue-side edge of the red peak, due to the infall motion of neutral gas around a galaxy, as seen in our Ly$\alpha$ spectrum. The cutoff location is found at $\sim 200$ km s$^{-1}$ in our spectra, which is comparable to the predicted velocity of the infalling gas around an $M_{\text{UV}} = -22$ galaxy at this redshift (Park et al. 2021).

Our Ly$\alpha$ line flux is $\sim$seven times fainter than the line flux measured from MOSFIRE. This could be due, in part, to the MOSFIRE slits being wider than the NIRSpec micro shutters (0$\farcs$7 versus 0$\farcs$2). Our NIRSpec slit-loss correction would not correct for extended Ly$\alpha$ emission in this source. The MOSFIRE observations of this source were taken $\sim$8–9 yr prior to the JWST/NIRSpec data, $\sim$1 yr in the rest frame, and it is possible that the variability common in AGNs might contribute to the difference in the measured line fluxes. AGN variability is typically $\sim$20% (e.g., MacLeod et al. 2012) and is unlikely to fully explain this large difference.

The N V$_{1243}$ line was detected from this source with Keck/MOSFIRE by Mainali et al. (2018) at 12019.5 Å, with the other line in the doublet, N V$_{1238}$, obscured by a skyline at 11981 Å. In our JWST/NIRSpec data, we do not detect a line at either of these wavelengths, but there is a 1.8$\sigma$ peak at 12007 Å. This would correspond to a rest-frame wavelength of 1240 Å between the peaks of the N V doublet. The velocity offset of the N V line from the systemic redshift $z_{\text{[O III]}} = 8.679$, if indeed the $\lambda$1238 line is $\sim +442 \pm 85$ km s$^{-1}$. This is consistent (within the error) with the offset we measure for the

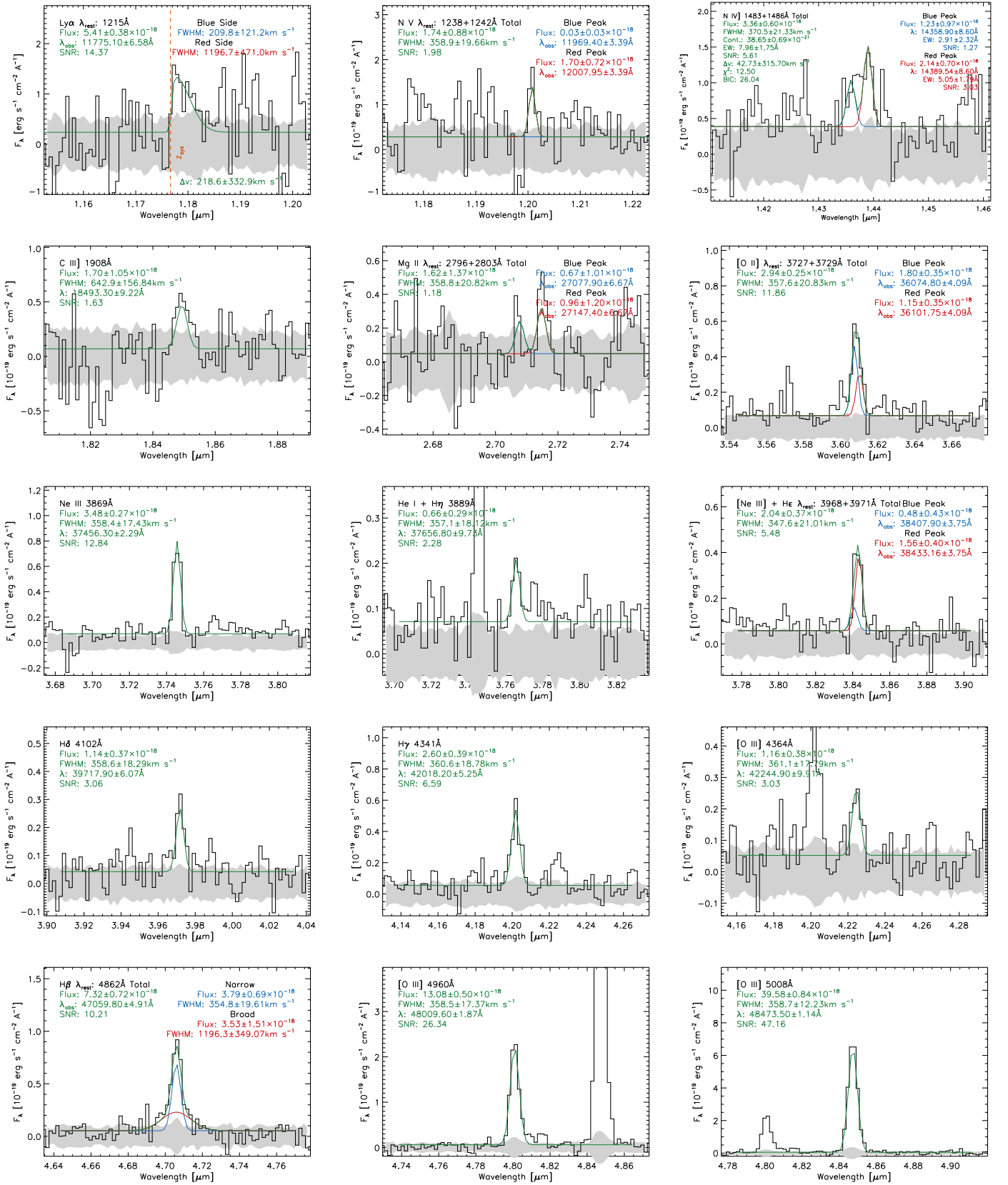

**Figure 5.** Fits to the emission lines identified in the JWST/NIRSpec spectrum of CEERS_1019. Each panel shows an individual emission-line fit, with the type of line profile as described in Section 3. The panels are plotted in $F_\lambda$ ($10^{-19}$ erg s$^{-1}$ cm$^{-2}$ Å$^{-1}$) vs. observed wavelength ($\mu$m), are presented in order of increasing wavelength, and are scaled vertically to show the extent of the highlighted emission line. Emission-line values are tabulated in Table 3 and discussions of specific lines are given in Section 4.

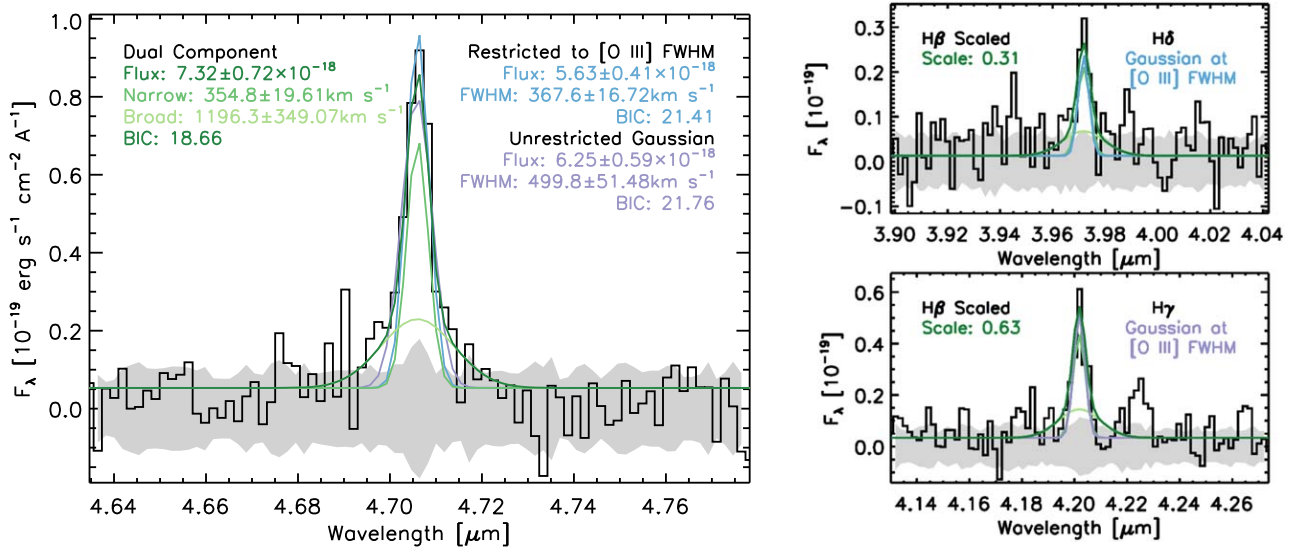

**Figure 6.** Left: dual-component Gaussian fit to the Hβ emission line (green), yielding FWHM$_{broad}$ = 1196.3 ± 349.1 km s$^{-1}$ and FWHM$_{nar}$ = 354.8 ± 19.6 km s$^{-1}$. Alternative single-Gaussian line profile fits are shown where the FWHM is restricted to FWHM $_{[O\,III]}$ ± 30 km s$^{-1}$ (blue) or left unrestricted (purple). In both cases, the fit to the Hβ emission line is worse than the dual-component fits, as indicated by the higher Bayesian Information Criterion (BIC) measurement (Section 4.2). Right: scaled-down fits of the same dual-component fit from Hβ to the peak of the Hδ (top) and Hγ (bottom) emission lines. In both cases, the broad component is not distinguishable from the noise, indicating that a broad component could be present in these lines below our noise threshold.

C III] of ∼ +310 ± 150 km s$^{-1}$. If it is the λ1243 line, the velocity offset is ∼ −521 ± 85 km s$^{-1}$.

### 4.4. Other Potential Broad Emission Lines

As noted above, broadline (Type 1) AGNs typically exhibit broad permitted lines and narrow forbidden lines, interpreted as high-velocity dense gas in a "broadline region" near the BH (and dominated by its gravitational influence) and lower-velocity gas in a "narrowline region" that is more distant from the BH and dominated by the galaxy kinematics (e.g., Netzer 2015). Unfortunately, the Hα line is redshifted beyond the wavelength range of NIRSpec for our target (though it is observable by MIRI spectroscopy). The other lines in the Balmer series (Hγ, Hδ, etc.) are too weak for strong constraints on the presence of broad components (see Figure 6). In AGNs with a broad Hβ line, broad features are also typically observed in the permitted UV lines Lyα, C IV, C III], and Mg II with similar or stronger fluxes to the broad Hβ line (e.g., Vanden Berk et al. 2001).

We detect a significant broad width in the observed Lyα line of our target. The blue-side absorption of the Lyα forest makes it difficult to measure the line width precisely, but we note that our asymmetric Gaussian fit (described in Section 3.3) finds a red-side width of FWHM = 1360.6 ± 479.9 km s$^{-1}$: comparable to the best-fit width of the broad Hβ component.

The best-fit Gaussian to the C III] line also indicates a broadline width of FWHM = 643 ± 157 km s$^{-1}$. The observed C III] feature is not well fit by two narrow Gaussians, with widths constrained to be the same as the [O III]λ5008 line for each of the doublet lines. This could indicate that the doublet is unresolved in a noisy part of our data, leading to a single line with a broader width. Alternatively, the broader FHWM of this line compared to [O III] may have some physical implications for this galaxy. The line width of the best-fit broad C III] profile is somewhat narrower than would be expected for a stratified broadline region, in which the higher-ionization C III] gas orbits closer to the BH and consequently has a broader velocity width than Hβ. The line center of the best-fit broad C III] profile is redshifted by +310 km s$^{-1}$ with respect to the systemic redshift (determined from the narrow [O III]λ5008 line as described in Section 4.1). The shift in line center and narrower-than-expected width may indicate a blueshifted absorption component in the C III] line, as commonly observed in broad absorption-line quasars (e.g., Gibson et al. 2009).

We additionally see a significant broad component in the best-fit profile for the N IV]λ1486 line. This feature is not typically observed in luminous quasars at $z \lesssim 4$, but it is detected as a feature in the NIRSpec prism spectrum of the $z = 10.6$ galaxy GN-z11 (Bunker et al. 2023). Its presence in GN-z11 has been interpreted as evidence for high-density and nitrogen-enhanced gas (Cameron et al. 2023; Senchyna et al. 2023), both properties consistent with expectations for broadline regions around rapidly accreting and low-mass AGNs (Hamann & Ferland 1999; Matsuoka et al. 2017).

On the other hand, we do not observe broad components in the C IV and Mg II emission lines. The absence of C IV can be explained by the low S/N in its region of the NIRSpec spectrum, and the large flux uncertainty at that wavelength range allows for an undetected broad C IV line that is of comparable brightness to the broad Hβ feature. The Mg II feature is more puzzling: the observed Mg II profile is best fit by narrow Gaussians for each of the doublet lines, consistent with galaxy (or narrowline region) emission, rather than the expected broad feature of an AGN. It is possible that this line is intrinsically broad but is affected by intervening absorption by the Mg II doublet: this requires a particular blueshift velocity of the absorption component, but is not uncommon in quasars. Alternatively, the AGN may be attenuated by dust, such that its UV emission lines are weaker than the Balmer lines, as implied by the best-fit SED model, although this conflicts with the detection of broad Lyα, N IV], and C III]. Deeper spectra of this source would more effectively test for the presence of the broad UV lines expected for AGNs, especially in investigating the hypothesis for blueshifted absorption affecting the broad C III] line and the Mg II line.

### 4.5. Emission-line Flux Ratio Measurements

Despite potential lingering issues in the absolute flux calibration of the NIRSpec instrument, we find that the relative flux calibration is consistent, at least for pairs of emission lines near one another in wavelength. We measure $[O\,III]_{5008}/[O\,III]_{4960} = 3.03 \pm 0.13$, which is consistent with the atomic physics calculation within errors (Storey & Zeippen 2000). We are confident in using the ratios of emission lines close in wavelength. Still, we acknowledge that additional flux calibration and improvements of the NIRSpec instrument are needed to trust widely separated line ratios fully. We also note that our wavelength-dependent aperture correction using the differing spatial profile of CEERS_1019 across the three NIRSpec filters may add additional systematic errors in distant line ratios. We thus only report ratios for lines that fall within the same NIRSpec filter: G395M. Since we also use the NIRCam filters to perform a wavelength-dependent flux correction (see Section 2.1.1), the errors on the line ratios that span across the two redder NIRCam filters (F356W and F444W) may be underrepresented. We report the following line ratios in this paper given the above caveats.

We list the measured line ratios for this source in Table 4. All reported line ratios are measured without using the broadline fit to the H$\beta$ line, and only the narrow component from the dual-component fit (See Figure 6). For any of the line ratios reported where one line is not detected at a significant level, we report the 1$\sigma$ upper limit for the line and mark it with an asterisk in the tables of line fluxes and ratios (Tables 3 and 4).

## 5. Constraints from Continuum Emission

We explore fitting the photometric SED of this object with stellar and AGN emissions simultaneously to explore which dominates the observed continuum emission. We first use the FAST v1.1 (Kriek et al. 2009; Aird et al. 2018) SED fitting code, including a star-forming galaxy and an AGN component, as described in Kocevski et al. (2023). With this code, we try fits using two sets of AGN templates. The first uses eight empirically determined AGN templates. These include five AGN-dominated templates from the Polletta et al. (2007) SWIRE template library (namely, the Torus, TQSO1, BQSO1, QSO1, and QSO2 templates) and three composite SEDs of X-ray-selected AGNs with absorption column densities of $N_H = 10^{22-23}$, $10^{23-24}$, and $10^{24-25}$ cm$^{-2}$ from Silva et al. (2004). See Appendix A of Aird et al. (2018) for additional details. We also try a second fit with a low-metallicity ($Z \leqslant 0.4 Z_\odot$) AGN component from CLOUDY modeling, chosen to mimic the SED of a radio-quiet AGN (see Section 6.5 for more details). We measure the ratio of light from the best-fitting AGN model to the total stellar+AGN model in two wavelength windows: a rest-UV window at 0.15–0.25 $\mu$m and a rest-optical window at 0.51–0.6 $\mu$m (both designed to avoid strong emission lines). In both of these fits, the photometry constrains the AGN to be subdominant to the stellar emission. With the fiducial AGN template, the stellar emission comprises 99.5% of the UV flux and 85% of the optical flux. With the low-metallicity AGN model, stellar emission comprises 82%–83% of both the UV and optical flux.

We try an alternative fit using the Cigale code (Boquien et al. 2019; Yang et al. 2020, 2022). In the fit, we adopt a modified Schartmann et al. (2005) AGN accretion disk model. We allow the deviation from the default optical spectral slope (the $\delta_{AGN}$ parameter) to vary from $-1$ to 1 and the polar dust extinction (the $E(B-V)_{PD}$ parameter) to vary from 0 to 1 (see Section 4 of Yang et al. 2022 for details). Measuring the results in the same two wavelength windows as above shows that the stellar emission comprises 70% and 59% of the continuum emission in the rest-UV and rest-optical, respectively.

These results imply that the observed continuum emission is dominated by stellar light. We thus proceed to perform Bayesian SED modeling with stellar-only models to explore what constraints can be placed on the stellar population of the host galaxy. Our fiducial results come from the Prospector SED fitting code (Johnson et al. 2021), following Tacchella et al. (2022) by using both a continuity and bursty prior on the star formation history (SFH). Both SFH priors lead to similar posterior distributions, implying that the data are informative and the impact of the prior is minimal. We find that this galaxy has a stellar mass of $M_\star \approx 10^{9.5 \pm 0.3} M_\odot$ and is actively forming stars (specific star formation rate, or sSFR $\approx 10^{-7.9 \pm 0.3}$ yr$^{-1}$), doubling its mass every $\sim$100 Myr. We find a mass-weighted age of $t_{50} = 34^{+119}_{-29}$ Myr. We also note that using the conversion from the star formation rate (SFR) to Balmer line flux from Kennicutt & Evans (2012), the H$\beta$ flux estimated from this SFR ($1.3-5.3 \times 10^{-18}$ erg s$^{-1}$ cm$^{-2}$) is fully consistent with our measured narrowline H$\beta$ flux ($3.8 \pm 0.7 \times 10^{-18}$ erg s$^{-1}$ cm$^{-2}$) within errors. We list the physical properties of this galaxy from our SED fitting in Table 5 and show the model fits in Figure 7.

We also fit these data with BAGPIPES (Carnall et al. 2018). We generally followed the procedures in Papovich et al. (2022), but we have now used Binary Population and Spectral

**Table 4**
Emission-line Ratios for CEERS_1019

| Emission-line Ratio | Measured Value | Convention |
|---|---|---|
| Oxygen | | |
| $[O\,III]_{5008}$ / $[O\,III]_{4960}$ | $3.03 \pm 0.13$ | |
| $[O\,III]_{4363}$ / $[O\,III]_{5008+4960}$ | $0.02 \pm 0.01$ | RO3 |
| $[O\,III]_{5008}$ / $[O\,II]_{3727+3729}$ | $13.46 \pm 1.18$ | O32 |
| O/H | | |
| $[O\,III]_{5008+4960}$ + $[O\,II]_{3727+3729}$ / H$\beta$ | $14.67 \pm 2.68$ | R23 |
| $[O\,II]_{3727+3729}$ / H$\beta$ | $0.78 \pm 0.16$ | |
| $[O\,II]_{3727+3729}$ / H$\gamma$ | $1.13 \pm 1.14$ | |
| $[O\,III]_{4363}$ / H$\gamma$ | $0.45 \pm 0.16$ | |
| $[O\,III]_{5008}$ / H$\beta$ | $10.44 \pm 1.91$ | R3 |
| Ne | | |
| $[Ne\,III]_{3869}$ / $[O\,II]_{3727+3729}$ | $1.18 \pm 0.14$ | |
| $[Ne\,III]_{3869}$ / H$\gamma$ | $1.34 \pm 0.22$ | |
| Balmer Series | | |
| H$\gamma$ / H$\beta$ | $0.69 \pm 0.16$ | |
| | $0.36 \pm 0.06$ | Total H$\beta$ |
| H$\beta$ / H$\gamma$ | $1.46 \pm 0.34$ | |
| | $2.82 \pm 0.49$ | Total H$\beta$ |
| H$\beta$ / H$\delta$ | $3.33 \pm 1.24$ | |
| | $6.42 \pm 2.18$ | Total H$\beta$ |
| H$\beta$ / H$\epsilon$ | $2.43 \pm 0.76$ | |
| | $4.69 \pm 1.29$ | Total H$\beta$ |

**Note.** Common emission-line flux ratios as measured from JWST/NIRSpec for CEERS_1019. All of the ratios that include H$\beta$ use the narrow component flux from the dual-component Gaussian fit (Section 3.4.3), as reported in column 6 of Table 4, unless otherwise noted as using the Total H$\beta$, which is the value in column 4.

**Table 5**
Physical Properties of CEERS_1019

| SFH Prior | $M_{\rm UV,obs}$ [mag] | UV Slope $\beta$ | log $M_\star$ [$M_\odot$] | log SFR$_{50}$ [$M_\odot$ yr$^{-1}$] | log sSFR$_{50}$ [yr$^{-1}$] | $A_V$ [mag] | log $Z$ [$Z_\odot$] |
|---|---|---|---|---|---|---|---|
| (1) | (2) | (3) | (4) | (5) | (6) | (7) | (8) |
| Continuous | $-22.07^{+0.05}_{-0.05}$ | $-1.77^{+0.11}_{-0.12}$ | $9.4^{+0.3}_{-0.2}$ | $1.5^{+0.2}_{-0.2}$ | $-7.9^{+0.2}_{-0.4}$ | $0.4^{+0.2}_{-0.2}$ | $-0.3^{+0.3}_{-1.0}$ |
| Bursty | $-22.09^{+0.06}_{-0.05}$ | $-1.80^{+0.09}_{-0.10}$ | $9.5^{+0.2}_{-0.4}$ | $1.5^{+0.4}_{-0.1}$ | $-7.9^{+0.2}_{-0.4}$ | $0.4^{+0.4}_{-0.2}$ | $-0.6^{+0.6}_{-1.1}$ |

**Note.** Physical properties of CEERS_1019 as derived through SED fitting to the HST and JWST photometry at the spectroscopic redshift. SED fitting with Prospector (Johnson et al. 2021) using both a continuity and bursty SFH prior, as described in Tacchella et al. (2022). Column (1); SFH prior used in the fit. Column (2): absolute UV magnitude at rest-frame 1500 Å. Column (3): UV spectral slope, $\beta$, as measured from the model spectra. Column (4): stellar mass. Column (5): SFR averaged over the past 50 Myr. Column (6): specific SFR averaged over the past 50 Myr. Column (7): dust attenuation at 5500 Å, from SED models. Column (8): stellar metallicity from SED models to the HST photometry. The values in columns (4) to (8) are calculated from SED fitting as in Tacchella et al. (2022).

Synthesis (BPASS) v2.2.1 stellar population models (Eldridge et al. 2017), an SFH represented as a Gaussian mixture model (Iyer et al. 2019), and fit over a range of ionization parameters log $U$ = [$-4$, $-1$] to model the nebular emission. From these fits, the galaxy has a stellar mass of log $M_\ast/M_\odot = 9.3 \pm 0.1$, consistent with the fits from Prospector, though a slightly higher log(sSFR) of $-7.5 \pm 0.2$ yr$^{-1}$. The systematic uncertainties on stellar mass and SFR here are $\approx$0.2 dex. The inferred properties, such as stellar mass, SFR, and dust attenuation, are generally consistent with and within the systematic uncertainties of the Prospector fits, but require a slightly higher SFR to account for differences in modeling assumptions. We show the best-fitting Prospector and BAGPIPES models in Figure 7, alongside images of this galaxy in all observed filters and photometric redshift results from EAZY.

### 5.1. UV Slope ($\beta$) from Photometry

To measure the UV spectral slope, $\beta$, for this source, we fit a power-law function ($f_\lambda \propto \lambda^\beta$; Calzetti et al. 1994) to the observed photometry. We only include filters within the rest-frame 1500–3000 Å range to avoid contamination from the Ly$\alpha$ break and strong emission lines. For this source, these filters are F160W, F150W, F200W, and F277W. Using the EMCEE software (Foreman-Mackey et al. 2013), we measure the posterior distribution on $\beta$. CEERS_1019 has $\beta = -1.76^{+0.12}_{-0.13}$, where the uncertainty is taken as the 68% central width from the posterior. This is consistent at the $\sim 1\sigma$ level with the value inferred from HST photometry from Tacchella et al. (2022) of $\beta = -1.61^{+0.18}_{-0.12}$. It is also consistent with the value of $\beta$ measured from the Prospector model spectra, listed in Table 5. This value of $\beta$ is comparable to the UV colors from other similarly bright sources (Tacchella et al. 2022), consistent with the low but non-negligible dust attenuation we find from our SED fit ($A_V = 0.4 \pm 0.2$).

### 5.2. Morphology

We investigate the morphology of this source using the fitting codes Galfit (Peng et al. 2002, 2010) and statmorph (Rodriguez-Gomez et al. 2019). We use the Galfit least-squares fitting algorithm to fit the galaxy's surface brightness profile. We use a 100 × 100 pixel cutout of the F200W science image as input, the corresponding error array (the "ERR" extension) as the input sigma image, and the empirically derived PSFs. As an initial guess, we use the source location, magnitude, size, position angle, and axis ratios from the Source Extractor (SE) catalog. The Sérsic index is allowed to vary between 0.01 and 8, the magnitude of the galaxy between 0 and 45, the effective radius ($R_e$) between 0.3 and 100 pixels, and the axis ratio between 0.0001 and 1. We also allow Galfit to oversample the PSF by a factor 9. We then visually inspect the best-fit model and image residual for each source to ensure that the fits are reasonable and that minimal flux remained in the residual.

We find that a single Sérsic component poorly fits the source. Visually, the source is extended asymmetrically, with three distinct components (see Figure 8). The optimum fit is obtained when the central region is fit with both a point source and a Sérsic component, and the two other regions to the west and the northeast are each fitted with their own Sérsic components. The requirement of a point-source component for the fit is consistent with the source having an AGN. We similarly performed fits on the F356W and F444W images and again found that single-Sérsic fits do not work because of the multiple components, even with the lower resolution at these longer wavelengths. The asymmetric nature of the source and the presence of the three separate components are consistent with the galaxy being involved in a major merger. We use statmorph to measure the concentration parameter and size as a function of wavelength on PSF-matched cutouts and find that C = 2.82 and $R_e$ = 6.49 pixels (0.91 kpc) for F356W and C = 2.93 and $R_e$ = 5.64 pixels (0.79 kpc) for F444W, suggesting that the F444W light is more concentrated than the F356W or F200W emission, supporting the visual impression in Figure 2. At the redshift of this galaxy, the F444W emission is dominated by the [O III] emission line (see Figure 7).

## 6. Constraints from Nebular Line Emission

### 6.1. Black Hole Mass

In this section, we estimate the virial mass of the central BH, assuming that the observed broad H$\beta$ emission traces the kinematics of gas in the broadline region. The measurement of BH masses with single-epoch spectra can be made using the width of the broad H$\beta$ emission line and the rest-frame 5100 Å continuum luminosity, $L_{5100}$, which has been shown to correlate with the distance to the broadline region using reverberation mapping (e.g., Kaspi et al. 2000; Cackett et al. 2021). However, this assumes the rest-frame 5100 Å continuum luminosity is dominated by light from the AGN, which is likely not the case here (see Section 6.1). As a result, we instead use Equation (10) from Greene & Ho (2005), which

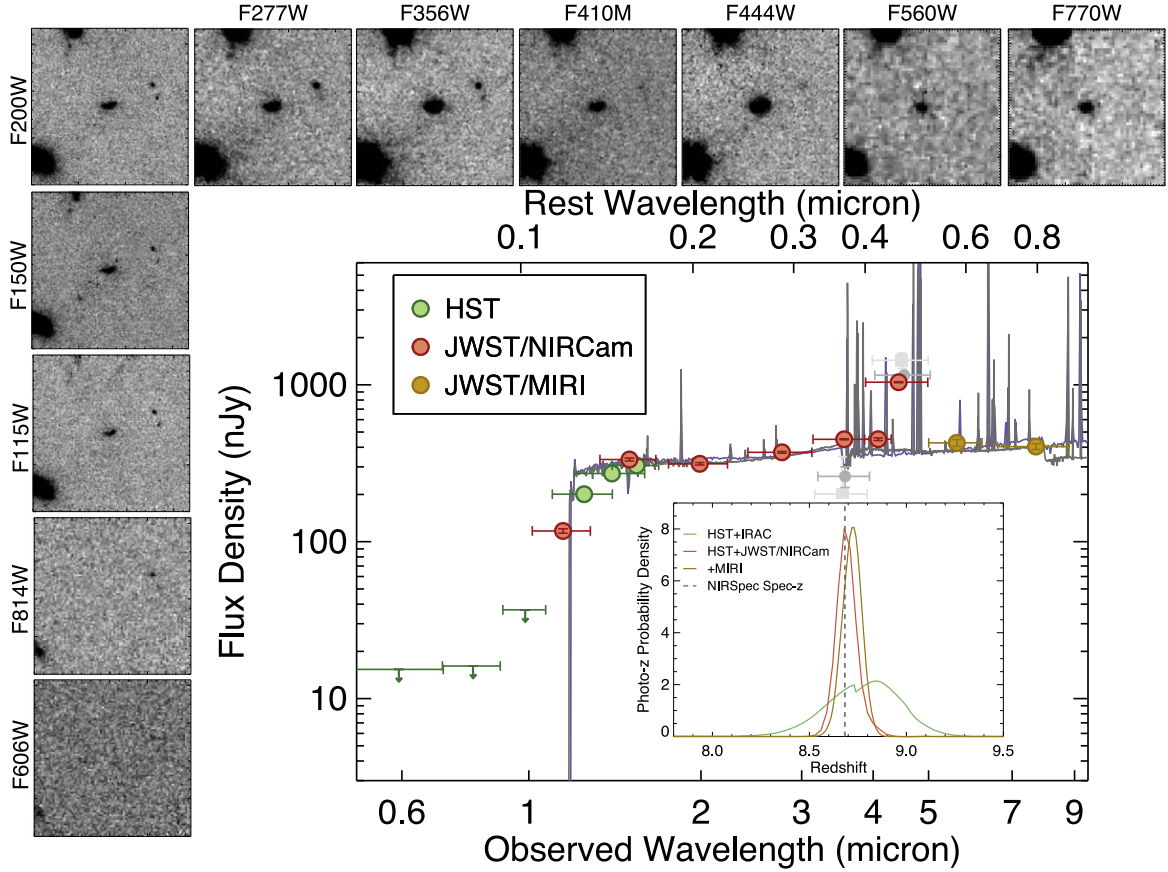

**Figure 7.** The images show 5″×5″ cutouts around this source in the CANDELS HST/ACS optical bands, CEERS NIRCam near-infrared bands, and CEERS MIRI bands. The source is a clear dropout in the optical and is well detected at 1–8 μm. The large inset plot shows the photometric SED from HST and JWST. We also show Spitzer/IRAC measurements from Finkelstein et al. (2022a) with both TPHOT (light gray) and Galfit (dark gray); this source was highly blended in IRAC, and the Galfit measurements appear closer to the NIRCam measurements. The lines are best-fit models from Prospector (gray) and BAGPIPES (purple). The SED is dominated by stellar emission (Section 5), and these stellar models infer a massive (log $M/M_\odot \sim 9.5$) and highly star-forming (log sSFR $\sim -7.9$ Gyr$^{-1}$) stellar population. The small inset panel shows constraints on the photometric redshift before (green) and after (red and yellow) the inclusion of JWST data. All three are consistent with the spectroscopic redshift, with those including JWST data placing much tighter constraints.

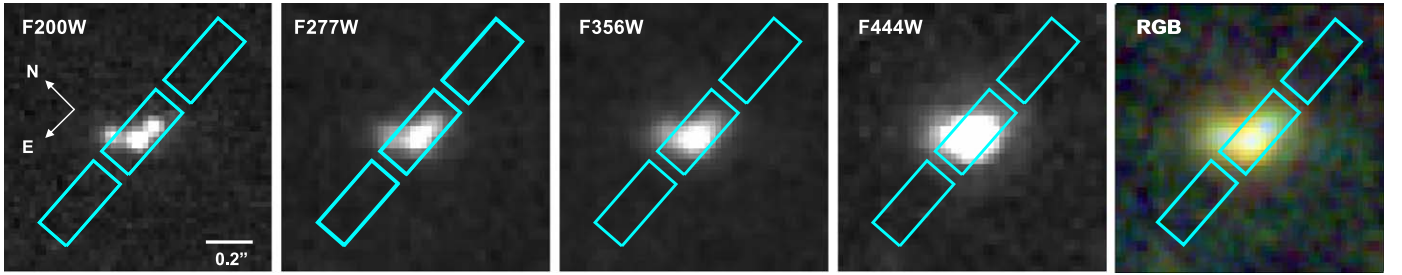

**Figure 8.** JWST NIRCam 1.5″ × 1.5″ cutouts of CEERS_1019 in four different filters (F200W, F277W, F356W, and F444W) and an RGB color composite (with blue = F115W + F150W + F200W, green = F277W + F356W + F410M, and red = F444W) with each filter at its native resolution, highlighting the substructure visible at shorter wavelengths. The positions of the NIRSpec MOS shutters are overlaid. This source has a bright central component centered in the shutter and two extended components as discussed in Section 5.2.

employs only the H$\beta$ line luminosity, $L_{H\beta}$, and width:

$$M_{BH} = 2.4 \times 10^6 \left(\frac{L_{H\beta}}{10^{42}\text{ erg s}^{-1}}\right)^{0.59} \left(\frac{\text{FWHM}_{H\beta}}{10^3 \text{ km s}^{-1}}\right)^2 M_\odot. \quad (5)$$

This equation is based on the formula of Kaspi et al. (2000), but with $L_{H\beta}$ substituted for $L_{5100}$ using the empirical correlation between Balmer emission-line luminosities and $L_{5100}$ reported in Greene & Ho (2005).

Using the luminosity and width of the broad H$\beta$ component, we derive a BH mass of $\log(M_{BH}/M_\odot) = 6.95 \pm 0.37$. This mass is 1–2 dex lower than existing samples of luminous quasars with BH mass estimates at $z > 5$ and more comparable to the low-luminosity AGN reported in Kocevski et al. (2023), found in these same CEERS NIRSpec data. Our measured mass implies that the AGN in CEERS_1019 is powered by the least massive BH known in the Universe during the EoR.

To determine the accretion rate onto the BH relative to the Eddington limit, we derive a bolometric luminosity, $L_{Bol}$, from the broad H$\beta$ line luminosity and compare it to the Eddington luminosity, $L_{Edd}$, for our measured BH mass. Assuming an intrinsic broadline H$\alpha$/H$\beta$ ratio of 3.06 (Dong et al. 2008) and

**Table 6**
CEERS_1019 BH Properties

| | | | |
|---|---|---|---|
| log $M_{BH}$ | 6.95 ± 0.37 | [M$_\odot$] | (1) |
| FWHM$_{H\beta, Broad}$ | 1196.26 ± 349.07 | [km s$^{-1}$] | (2) |
| log $L_{H\beta, Broad}$ | 42.5 ± 0.2 | [erg s$^{-1}$] | (3) |
| log $L_{Bol}$ | 45.1 ± 0.2 | [erg s$^{-1}$] | (4) |
| $\lambda_{Edd}$ | 1.3 ± 0.5 | | (5) |

**Note.** Column (1): BH mass (Section 6.1). Column (2): FWHM of the broad H$\beta$ line (Section 4.2). Column (3): luminosity of the broad H$\beta$ line. Column (4): bolometric luminosity. Column (5): Eddington ratio ($L_{Bol}/L_{Edd}$).

a bolometric correction of $L_{Bol} = 130 \times L_{H\alpha}$ (Richards et al. 2006; Stern & Laor 2012), we obtain $L_{Bol} = 1.4 \pm 0.6 \times 10^{45}$ erg s$^{-1}$. This results in an Eddington ratio, $\lambda_{Edd} = L_{Bol}/L_{Edd}$, of 1.3 ± 0.5. This suggests that the BH is undergoing a rapid growth phase and is accreting at approximately its Eddington limit.

We note that while we observe a broad H$\beta$ line, we do not detect AGN continuum emission at the level that might be expected given the Greene & Ho (2005) relationship between $L_{H\beta}$ and $L_{5100}$. This may indicate the AGN is moderately reddened due to a patchy obscuring medium, similar to what is found in red quasars, which show broad Balmer lines, yet have properties intermediate between type I and II quasars (Greene et al. 2014; Glikman et al. 2023). If this is the case and the H$\beta$ emission is suppressed by dust extinction, then our estimated BH mass should be considered a lower limit to the true virial mass for this AGN. We report our measurements fo the BH in CEERS_1019 in Table 6.

### 6.2. Electron Temperature and $T_e$-based Metallicity

The [O III]$_{4364}$/ [O III]$_{5008+4960}$ ratio can be used to measure the electron temperature ($T_e$) of the galaxy's interstellar medium (ISM). As these lines are all collisionally excited, a higher [O III]$_{4364}$ emission relative to [O III]$_{5008+4960}$ indicates higher-energy electrons are responsible for the excitation. This $T_e$ has been found to correlate with the ISM metallicity, providing a way to measure the metallicity of the source with these lines (i.e., Kewley et al. 2019b).

Our source has [O III]$_{4364}$/ [O III]$_{4960+5008}$ = 0.022 ± 0.007. We use the measured [O III]$_{4364}$/ [O III]$_{4960+5008}$ ratio to estimate $T_e$ using Equation (4) of Nicholls et al. (2020):

$$\log_{10}(T_e) = \frac{3.5363 + 7.2939x}{1.0000 + 1.6298x - 0.1221x^2 - 0.0074x^3}, \quad (6)$$

where

$$x = \log_{10}\left(\frac{[O\ III]_{4364}}{[O\ III]_{4960+5008}}\right) = -1.657 \pm 0.003.$$

This yields $\log_{10}(T_e) = 4.270 \pm 0.566$ and $T_e = 18630.755 \pm 3.682 K$ for CEERS_1019.

By measuring $T_e$ in this way, we are only sensitive to the portion of the ISM emitting the [O III] lines, which may not represent the entire galaxy. Any significant gradients in density and/or ionization may lead to a mixture of different regions in the galaxy, and the $T_e$ from [O III] is only sensitive to the high-ionization regions.

We then use Equation (1) from Pérez-Montero et al. (2021) to estimate the metallicity from $T_e$ as shown below:

$$12 + \log(O/H) = 9.72 - 1.70t_e + 0.32t_e^2, \quad (7)$$

where $t_e = T_e$ in units of $10^4$K and thus our "direct" metallicity measurement for this source is: $12 + \log(O/H) = 7.664 \pm 0.508$ or $0.095^{+0.209}_{-0.065} Z_\odot$ ($Z_\odot = 8.69$; Asplund et al. 2021).

This method is used for extreme emission-line galaxies (EELGs), galaxies with extreme emission lines dominated by stellar photoionization, so it may not be strictly accurate for objects whose photoionization is dominated by an AGN. The above equation from Pérez-Montero et al. (2021) is adopted from Amorín et al. (2015), who derived this relation based upon calibrations using local EELGs, after discarding AGN candidates based upon "Baldwin–Phillips–Televich" (BPT; Baldwin et al. 1981) diagram measurements.

### 6.3. Electron Density from [O II] Doublet

The ratio between the two lines in the [O II]$_{3727+3729}$ doublet is often used to determine the electron density ($n_e$) in star-forming regions at temperatures T ~ 10,000 = 20,000 K. This is because the excitation energy between the lines is on the order of the thermal electron energy. Thus the relative excitation rates depend only upon collision strengths (Osterbrock 1989). While the [O II]$_{3727+3729}$ doublet is blended at the resolution of the NIRSpec data (R~1000 in the M gratings), the separation between the two line peaks at this redshift is large enough to distinguish. The doublet fit to this line (Section 3.4.2) gives flux values for both emission lines such that we can estimate a ratio between the two of [O II]$_{3729}$ / [O II]$_{3727}$ = 0.639 ± 0.255 with which to infer an electron density, $n_e = 1.9 \pm 0.2 \times 10^3$ cm$^{-3}$ (Osterbrock 1989). This is consistent with the feedback-free star formation model from Dekel et al. (2023). Alternatively, measurements using the C III] doublet are sensitive to larger $n_e$. We only detect the C III]$_{1908}$ line from this doublet, which could imply that CEERS_1019 has very high density, $\sim n_e > 10^4$ cm$^{-3}$ (Keenan et al. 1992).

### 6.4. A$_v$ from Balmer Decrement

The ratios of Balmer lines can be used to measure nebular attenuation. In particular, gas with Case B recombination and a temperature of $T_e = 20,000$ K (as calculated for our object in Section 6.2) will have intrinsic Balmer ratios of H$\beta$/H$\gamma$ = 2.110 and H$\beta$/H$\delta$ = 3.817 (Osterbrock 1989). We can estimate the dust attenuation E(B–V) with an assumed attenuation curve (here, we use Calzetti 2001), by comparing our observed line ratios to these intrinsic values. For the observed line ratios, we use the total line flux rather than the narrowline-only flux, because the H$\gamma$ and H$\delta$ lines do not have sufficient S/N to separate broad and narrow components. This means that our estimated nebular attenuation may be a mix of attenuation affecting the AGN and the narrow lines.

From the observed line ratios of H$\beta$/H$\gamma$ = 2.82 ± 0.49 and H$\beta$/H$\delta$ = 6.42 ± 2.18, we find nebular attenuation of:

1. $E(B-V)_{H\beta/H\gamma} = 0.64 \pm 0.41$;
2. $E(B-V)_{H\beta/H\delta} = 0.83 \pm 0.54$.

The observed line ratios are statistically consistent with the intrinsic values, such that the implied $E(B-V)$ values are similarly consistent with little to no dust. But the large error bars also allow for a broad range of nebular attenuation. In Section 5.1, the Prospector SED fit implied $A_V = 0.4 \pm 0.2$, which equates to $E(B - V) = 0.1 \pm 0.05$. This is consistent within the large error bars of the Balmer decrement estimate for dust attenuation.

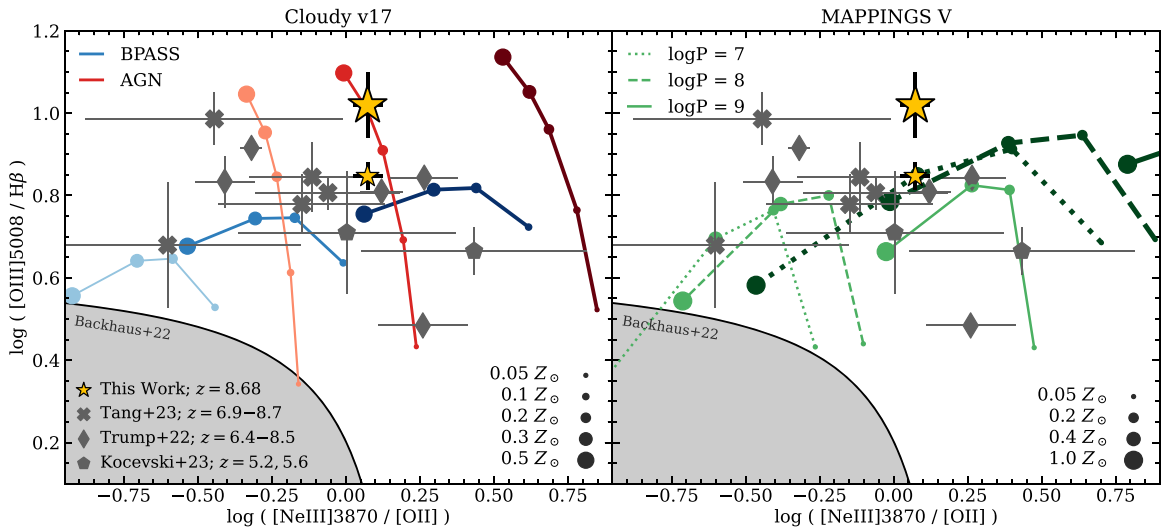

**Figure 9.** The "OHNO" diagram, the line ratio diagnostic using [O III]$_{5008}$/H$\beta$ vs. [Ne III]$_{3870}$/[O II]$_{3727+3729}$, with the black curve showing the boundary between star-forming and AGN regions at $z \sim 1$ (as defined in Backhaus et al. 2022, with the shaded region showing the star-forming area). The orange stars are from this work, representing the narrow component fit (big star) and the single-Gaussian fit (to the full profile; small star) of the H$\beta$ emission line in our galaxy. The black polygons represent other high-redshift ($z > 5$) galaxies with JWST/NIRSpec observations (Trump et al. 2023; Kocevski et al. 2023; Tang et al. 2023). Underplotted on both panels are photoionization models, with the left panel showing star-forming (blue) and AGN models (red) using the CLOUDY v17 code (Ferland et al. 2017) and the right panel showing star-forming models (green) from Kewley et al. (2019a) using the MAPPINGS V code (Sutherland et al. 2018). Each line of points represents a specific set of ionization parameters of $\log_{10} U = [-2.5, -2.1, -1.5]$ for the CLOUDY models and $\log_{10} U = [-2.5, -1.5]$ for the MAPPINGS models, with increasing ionization toward darker colors (or toward the top right of each panel). For the MAPPINGS models, the lines are further separated by isobaric pressure (denoted by line type). The sizes of the points represent the metallicities of the models.

### 6.5. Ionization Parameter

To[48] place this source in the context of other star-forming galaxies and AGNs identified in this early epoch, we investigate the "OHNO" line ratio diagnostic diagram. This diagram, comparing [O III]$_{5008}$/H$\beta$ versus [Ne III]$_{3870}$/[O II]$_{3727+3729}$, has been used at lower redshifts to identify ionization powered purely by star formation from that of an AGN (e.g., Backhaus et al. 2022; Cleri et al. 2023, 2023). Recent studies of high-redshift galaxies ($z > 5$) using JWST spectroscopy have found most if not all of their sources occupying the AGN region of the diagram (i.e., above the theoretical line dividing star formation from AGN, calibrated via low-redshift observations)—results that seem to suggest strong ionization from AGNs and/or very-metal-poor star-forming H II regions at these high redshifts (e.g., Trump et al. 2023; Kocevski et al. 2023; Tang et al. 2023; Übler et al. 2023).

Such is the case for CEERS_1019, with high measurements of the ionization indicators [O III]$_{5008}$/H$\beta$ ≡ R3 = 10.44 ± 1.91 and [O III]$_{4960+5008}$/[O II]$_{3727+3729}$ ≡ O32 = 13.46 ± 1.18 (see Table 4). Figure 9 shows our galaxy on the "OHNO" diagram (orange stars), using the narrow component of the H$\beta$ flux (the bigger star—also the light green line in Figure 6; the smaller star shows the ratio with the full H$\beta$ flux). Additional sources from recent JWST studies focused on the EoR (Trump et al. 2023; Kocevski et al. 2023; Tang et al. 2023) are included as small gray polygons. The black curve in both panels denotes the boundary between star-forming and AGN regions as defined by Backhaus et al. (2022), with the associated gray shaded region showing the $z \sim 1$ "star-forming" region. Underplotted are photoionization models from both the CLOUDY v17 (Ferland et al. 2017; left panel) and MAPPINGS V (Sutherland et al. 2018; right panel) codes, with parameters chosen to showcase a range of possible galaxy properties at this redshift, including elemental abundances, stellar populations, and stellar and gas-phase metallicities. In this figure, each line in the left panel represents CLOUDY models with a specific ionization parameter (increasing $\log_{10} U$, shown by the darker colors), where the sizes of the points in each line represent the specific gas-phase metallicity of said model. Similarly, each line in the right panel represents the MAPPINGS models for a specific ionization parameter (similar to the left panel) and isobaric pressure (denoted by the line types).

The CLOUDY models shown in Figure 9 assume a plane-parallel geometry with a nebular electron density of $10^3$ cm$^{-3}$ (inferred from the flux ratio of the [O II]$_{3727+3729}$ doublet; see Section 6.3), scaled Solar elemental abundances, and cover ionization parameters from $\log_{10} U = [-2.5, -2.1, -1.5]$. The star-forming models use BPASS v2.0 (Eldridge et al. 2017) fiducial binary stellar population models with an IMF that extends to 300 M$_\odot$ (described by a slope of $\alpha = -1.30$ from 0.1 to 0.5 M$_\odot$ and $\alpha = -2.35$ from 0.5 to 300 M$_\odot$) and continuous star formation of 1 M$_\odot$ yr$^{-1}$. For these models, we fix the stellar and gas-phase metallicities, covering $Z = 0.1$–0.5 $Z_\odot$. The AGN models use the `table agn` model in CLOUDY, which approximates a "typical" radio-quiet AGN, covering the same range of ionization parameters as the star-forming models and spanning nebular metallicities of 0.05–0.5 $Z_\odot$.

The MAPPINGS models shown in the same figure are the "Pressure Models" from Kewley et al. (2019a), assuming a plane-parallel geometry with a range of pressure (log$_{10}(P/k)=7$–9 cm$^{-3}$), ionization parameter (log$_{10}U = [-2.5, -1.5]$; also referred to as log$_{10}Q = 8$–9), and metallicity ($Z = 0.05 - 1\ Z_\odot$). These star-forming models use the Starburst99 (Leitherer et al. 2014) stellar population models with a Salpeter IMF (Salpeter 1955) extending to 100 M$_\odot$. The stellar and gas-phase metallicities follow a prescription such that at lower metallicities, the models have increasingly enhanced $\alpha$ abundances (see Nicholls et al. 2017 for more details). We have included these models in addition to the CLOUDY models (which have Solar abundances scaled to chosen metallicity) to showcase the range of potential galaxy properties for our high-redshift galaxy.

---

[48] Defined as the Ratio of the Number Density of Ionizing Photons to the Gas Density, $U \equiv n_\gamma / n_H$

Similar to that found in other spectroscopic studies in this epoch (e.g., Trump et al. 2023; Kocevski et al. 2023; Tang et al. 2023), the line ratio when using the single-component fit to the H$\beta$ line for CEERS_1019 (the smaller star) lies in a region on this diagnostic that is difficult to differentiate between AGN and metal-poor high-ionization star-forming H II regions. This is not unexpected, as galaxies at higher redshifts have been shown to generally have higher ionization and lower metallicities in comparison to those at lower redshifts (e.g., Shapley et al. 2003; Erb et al. 2010; Kewley et al. 2019a; Backhaus et al. 2022; Papovich et al. 2022; Sanders et al. 2023). This suggests that at these high redshifts, successfully separating star-forming and AGN sources using these strong line diagnostics can be challenging—evidenced by the lower-metallicity AGN models overlapping with the star-forming models in the left panel of Figure 9. Further discussion about the general utility of such emission-line diagnostics at high redshift is discussed in Section 7.3.

However, when focusing on the fit to only the narrow component of the H$\beta$ line for CEERS_1019 (the bigger star), there is a clear distinction from the rest of the high-redshift sources shown in this figure. Of the sources shown, our galaxy is the only one with enough S/N in H$\beta$ to (a) see a broadening of the line profile, and (b) clearly measure both a broad and narrow component for the line.[49] Follow-up spectroscopy of these and other sources that fall in this strong line regime at high redshift may shed light on more "hidden" AGNs in the early Universe (Kocevski et al. 2023).

From Figure 9, CEERS_1019 covers a similar location to the CLOUDY star-forming models with log$_{10}U \sim -1.9$ to $-1.5$ (for the single-component fit to H$\beta$; smaller star) and the AGN models with log$_{10}U \sim -2.1$ (for both the narrow and single-component fits to H$\beta$; both stars). Similarly, the MAPPINGS models suggest agreement with log$_{10}U \sim -2.5$ to $-1.5$; however, this is also dependent upon the pressure assumed. These results add to the expectation that this galaxy has high ionization powering these strong line ratios. Indeed, this agrees well with the strong O32 line ratio measured for this galaxy, which is often indicative of highly ionized, metal-poor gas (e.g., Schaerer et al. 2022; Williams et al. 2023) and a regime that could suggest Lyman continuum leakage (e.g., Izotov et al. 2018). As a comparison to the ionization parameters gleaned from Figure 9, we estimate this value using two relations from the literature. First, using the theoretical relation between O32, 12+log(O/H), and the ionization parameter described in Kobulnicky & Kewley (2004), we derive log$_{10} \sim -2.3$.

Finally, using the measured relation between O32 and the ionization parameter explored in Papovich et al. (2022), we derive log$_{10} \sim -2.0$. These results are all relatively consistent with one another and further highlight the strong ionizing nature of this source.

### 7. Discussion

The "chicken or egg" origin of BH seeds in the first galaxies remains unsolved. Theoretical predictions suggest a mix of "light" ($\sim 10^2$ M$_\odot$) Population III stellar remnants and/or "heavy" ($\sim 10^5$ M$_\odot$) seeds formed via the direct collapse of primordial gas, as summarized in recent reviews by Smith & Bromm (2019), Inayoshi et al. (2020), and Fan et al. (2022), but the relative mix of each seed type remains unconstrained by observations. Regardless of the seeding mechanisms, our Universe is capable of forming extremely massive SMBHs very early, with the highest-redshift massive quasar at the time of this writing being $z = 7.642$ (Wang et al. 2021), existing $\lesssim 700$ Myr after the Big Bang. Observations with JWST, especially MIRI observations of obscured growth (e.g., G. Yang et al. 2023, in preparation) and NIRSpec observations of broad ($\gtrsim 1000$ km s$^{-1}$) and/or high-ionization (e.g., N V, He II, [Ne V], and C IV) emission lines can now allow the first real census of AGNs in low-mass ($M_* < 10^{10}$ M$_\odot$) hosts at high redshift, providing strong constraints on the initial BH seed distribution. Direct observations of early SMBHs will not only constrain mechanisms for early BH growth (e.g., Ricarte & Natarajan 2018a, 2018b), but can also provide further insight into the role ionizing photons from AGNs played in the reionization of the IGM (e.g., Finkelstein et al. 2019; Giallongo et al. 2019; Dayal et al. 2020; Grazian et al. 2020; Yung et al. 2021; Grazian et al. 2022).

#### 7.1. Constraints on the Formation of This $z = 8.68$ AGN

Discovering an actively accreting SMBH at such a high redshift further constrains the formation time for such objects. Here we consider plausible formation mechanisms for this source (e.g., small versus large seeds), as well as discuss what its plausible descendants are. While the inferred mass of this BH is not much larger than the proposed range for DCBH seeds, it is unlikely that we are witnessing a DCBH soon after formation. The constraints we place on both the stellar mass and gas-phase metallicity of the host galaxy are significantly higher than expected (DCBH formation requires near-primordial conditions), implying that this object is a "standard" (albeit very distant) AGN accreting at $\sim$the Eddington rate, with a continuum SED dominated by stellar emission (in line with the stellar emission dominating the continuum SED; Section 5), similar to the scenarios recently explored by Volonteri et al. (2023).

It is therefore interesting to explore how this AGN was seeded: from a low-mass ($\sim$10–100 M$_\odot$) stellar seed or a high-mass ($\sim 10^{4-6}$ M$_\odot$) DCBH seed, summarized in Figure 10. In between these two scenarios could be a seed from a massive starburst cluster or young ultracompact dwarf galaxy (Kroupa et al. 2020) labeled as "Dense Star Cluster Seed" in Figure 10. The purple curves show idealized BH mass tracks assuming a 100 M$_\odot$ stellar seed. Such a seed could plausibly form at $z \sim 30$, but would be unable to accrete for $\sim$100 Myr due to radiative heating of the gas from the stellar progenitor (e.g., Johnson & Bromm 2007; Jeon et al. 2014). Assuming such a

---

[49] In Kocevski et al. (2023), while they could resolve both components for the H$\alpha$ emission in the two AGNs in their sample, the H$\beta$ emission for both sources had too low S/N to enable this identification.

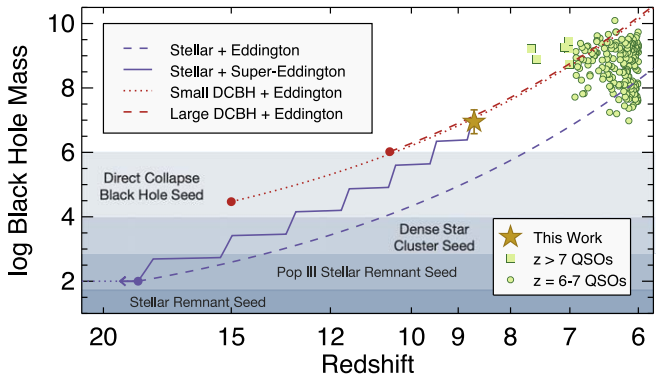

**Figure 10.** Mass of BHs vs. redshift. The gold star represents the result of this work. The green points denote published $z = 6$–7.64 quasars, taken from the compilation of Inayoshi et al. (2020) and augmented by more recent $z > 7$ quasars from Fan et al. (2022). The different lines show potential tracks of BH growth. The purple lines show a 100 $M_\odot$ Population III stellar remnant seed forming at $z \sim 30$, with growth beginning after a 100 Myr delay due to progenitor gas heating. The dashed purple line shows constant Eddington growth, which is both likely unphysical and cannot reproduce our observed object mass. The solid line shows one example of how periods of super-Eddington growth (10 times Eddington for 10 Myr) separated by periods of sub-Eddington growth (0.1 times Eddington for 50 Myr) could plausibly create the observed source. The red lines show potential formation mechanisms from a small ($3 \times 10^4\,M_\odot$) DCBH forming at $z = 15$ (dotted) and a large ($10^6\,M_\odot$) DCBH forming at $z = 10.5$ (dashed), each growing at the Eddington rate. Either of the mechanisms that could explain our observed BH source, DCBH+Eddington or stellar+super-Eddington, are somewhat exotic scenarios, pushing standard assumptions. Probing SMBHs out to higher redshifts and lower masses will clarify the formation mechanisms of these objects.

100 Myr delay, the onset of BH growth would begin by $z \sim 18.5$. As shown by the dashed purple line, Eddington-limited accretion in this scenario would be unable to reach the inferred BH mass by the redshift measured for this object. The enhanced stellar feedback environment the host galaxy must have experienced in the previous ∼100 Myr further complicates this. Building up to the observed $Z \gtrsim 0.01 Z_\odot$ requires a large number of supernova explosions in a short period of time. This would lead to violent, turbulent mixing and heating of the gas, making accretion likely very inefficient (significantly sub-Eddington).

However, there may also be periods of super-Eddington ("catch-up") accretion. These would occur in short episodes, with minimal sub-Eddington accretion between them; such short bursts of BH growth could plausibly build up the $\sim 10^7\,M_\odot$ SMBH we observe in an otherwise stellar-emission-dominated galaxy (e.g., Volonteri & Rees 2005; Madau et al. 2014; Inayoshi et al. 2016). For example, Pezzulli et al. (2016), when examining the evolution of merger tree simulations, find that super-Eddington accretion modes (e.g., Haiman 2004; Silk 2005; Polletta 2008) can be a significant component of SMBH growth in gas-rich environments, where up to 75% of the SMBH growth can be accounted for by periods of super-Eddington accretion, with intermittent phases of disruption resulting from the rapid depletion/replenishment of the bulge gas reservoir out of which the BHs accrete.

In Figure 10, we highlight one potential version of this scenario, with the purple solid line showing an object that starts with a seed mass of 100 $M_\odot$, which begins growing at $z = 18.5$ with episodic 10 Myr periods of super-Eddington (10 times) growth, followed by 50 Myr "breathing" periods with sub-Eddington (0.1 times) growth. Periodic super-/sub-Eddington growth allows this hypothetical object to grow to $\sim 10^7\,M_\odot$ by $z \sim 8.7$ from a small seed, matching our observations. While plausible (in particular, the episodic growth with suppressed, sub-Eddington, accretion periods was predicted by the simulations of Jeon et al. 2012 and Massonneau et al. 2023), this scenario is somewhat contrived, making a stellar seed somewhat unlikely.

A DCBH origin may be more plausible. While the inferred metallicity implies the DCBH event occurred significantly prior to the observed epoch, this could provide the needed time for a massive seed $\sim 10^6\,M_\odot$ to grow by the needed factor of $\sim 10$ times, without also having to invoke super-Eddington conditions. Given the likely strong stellar feedback environment in the assembling galaxy, this DCBH scenario may be favored. We show two plausible DCBH tracks as the red lines in Figure 10. Both options, of a lower-mass $3 \times 10^4\,M_\odot$ DCBH forming at $z = 15$ or a higher-mass $10^6\,M_\odot$ DCBH forming at $z = 10.5$, should they grow at the Eddington limit, could reproduce the observed mass of this object by $z \sim 8.7$.

Last, extrapolating the inferred BH growth tracks to lower redshifts provides insight into the ultimate descendants of early BHs, such as the one we observe here. As shown in Figure 10, this object is both too low-mass and observed too late to be the plausible progenitor of the $z > 7$ quasar population. However, should near-Eddington accretion be sustained for significant periods of time, objects like this one could plausibly evolve into the massive $z \sim 6 - 7$ SDSS quasar population. Further constraining both the progenitors and descendants of high-redshift SMBHs is possible, as the population of $z > 8$ SMBHs is coming into view. Given the discovery of this source in a relatively small data set, it is likely that further JWST spectroscopy—in particular, following up bright sources discovered over wide areas with the upcoming Nancy Grace Roman Space Telescope—will yield such a population (e.g., Yung et al. 2023).

### 7.2. Evolution in the BH–Galaxy Mass Relationship

Figure 11 compares CEERS_1019 with other high-redshift ($z \sim 6$) AGNs (Izumi et al. 2021; Kocevski et al. 2023) as well as the $z = 0$ $M_{\rm BH} - M_*$ relationship. Because they are easier to detect with pre-JWST instruments, most high-redshift AGNs are high luminosity. However, AGNs have a well-known luminosity-dependent bias (i.e., the Lauer bias; Lauer et al. 2007), such that higher-luminosity AGNs tend to have overmassive BHs.

JWST has improved our ability to select low-luminosity accreting BHs (e.g., Kocevski et al. 2023), which are expected to lie closer to the intrinsic (overall) $M_{\rm BH} - M_*$ relationship. Figure 11 shows an estimate from the empirical TRINITY model (Zhang et al. 2023a) for how different bolometric luminosity thresholds bias the median $M_{\rm BH} - M_*$ relationship. With an estimated bolometric luminosity of $L_{\rm bol} \sim 10^{45}$ erg/s and a host stellar mass of $M_* \sim 10^{9.5}\,M_\odot$, TRINITY would suggest that the present source would be expected to have a $\Delta M_{\rm BH} \sim 0.7$ dex offset from the median relation. The present source is located around the upper envelope of the $M_{\rm BH} - M_*$ relation spanned by the AGN sample from Reines & Volonteri (2015) at $z \sim 0$. The present source also has a lower $M_{\rm BH}$ compared to the $z = 5.55$ AGN from Übler et al. (2023), which is $\sim 10$ times brighter. Qualitatively, this is consistent with the Lauer bias, that brighter AGNs tend to be overmassive BHs compared to their host galaxies.

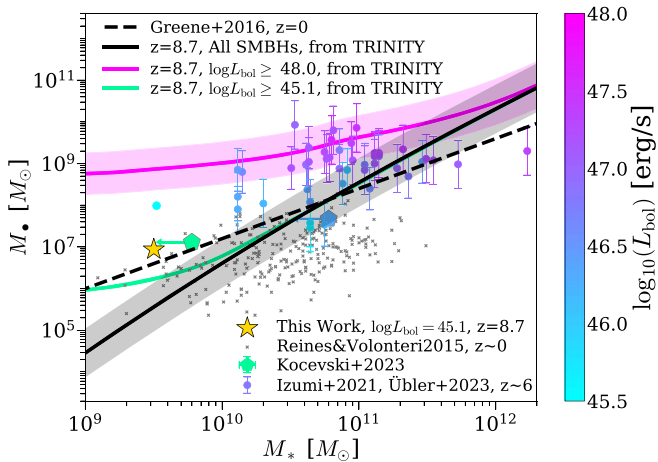

**Figure 11.** The predicted $z = 8.7$ median $M_\bullet - M_*$ relation for quasars with different bolometric luminosity thresholds from the TRINITY model (solid curves; Zhang et al. 2023a, 2023b). The pink shaded region is the $1 - \sigma$ spread around the median scaling relation for quasars (∼0.55 dex), which includes the random scatter in observed $M_\bullet$ when using virial estimates. This (log-)normal scatter is nearly luminosity-independent, so we only show it for the brightest quasars for clarity. The black solid line is the predicted $M_\bullet - M_*$ relation for all SMBHs at $z = 8.7$, and the black shaded region is the intrinsic+observed scatter around the intrinsic $M_\bullet - M_*$ relation. The green solid line is the $M_\bullet - M_*$ relation for AGNs brighter than $\log L_{\rm bol}[{\rm erg/s}] \geqslant 45.1$, corresponding to the bolometric luminosity of the AGN in this work (gold star). For comparison, we also show the $z = 0$ relation from Greene et al. (2016) with the black dashed line and the $z = 0$ AGN sample from Reines & Volonteri (2015) with the black crosses. The following data are shown in symbols color-coded by bolometric AGN luminosities: (1) $z \sim 6$ quasars compiled by Izumi et al. (2021) and Übler et al. (2023; filled circles); and (2) the two $z > 5$ AGNs from Kocevski et al. (2023; pentagons).

If the present source lies on the median $M_{\rm BH} - M_*$ for its luminosity, the Lauer bias estimate from TRINITY would suggest that the overall $z = 8.7$ median $M_{\rm BH} - M_*$ would have $M_{\rm BH} = 10^{6.3}~M_\odot$ at $M_* = 10^{9.5}~M_\odot$, i.e., there would be no overall evolution from the $z = 0$ $M_{\rm BH} - M_*$ relationship. If other sources are confirmed to have similarly consistent masses compared to the $z = 0$ $M_{\rm BH} - M_*$ relationship, it would cement a tight relationship between galaxies and BHs that extends into the EoR, and would also provide a strong challenge to theoretical models that have predicted substantial evolution in the $M_{\rm BH} - M_*$ relation for high-redshift and low-mass galaxies (see Habouzit et al. 2021 for a survey).

### 7.3. The Efficacy of Line Ratio Diagrams in Identifying AGNs at Early Times

Many of the canonical emission-line ratio diagnostics of AGN versus star formation using rest-frame optical emission lines are calibrated for low to moderate redshifts ($z \lesssim 1$; e.g., Baldwin et al. 1981; Veilleux & Osterbrock 1987; Juneau et al. 2011, 2014; Trump et al. 2015; Backhaus et al. 2022). As such, the divisions between the star-forming and AGN regions defined by these optical line ratio diagnostics are not necessarily valid at higher redshifts.

Several works offer remedies to these low-redshift diagnostics by quantifying a redshift evolution to the division (e.g., Coil et al. 2015; Cleri et al. 2023), but these new divisions have not been extended to the EoR; in fact, these relations for these diagrams are shown to break down at high $z$ in simulations (Hirschmann et al. 2019, 2022). The redshift-evolving diagnostics of these works employ a simple shift from the low-redshift divisions; however, there is insufficient knowledge of high-redshift AGNs to validate this practice for galaxies in the EoR.

For a different avenue to remediate these issues, other works have offered completely new emission-line ratio diagnostics with higher-ionization emission-line ratios, e.g., in the optical with He II/H$\beta$ and [Ne V]/[Ne III] (Katz et al. 2023; Cleri et al. 2023) and in the UV with, e.g., C III]/He II, O III]/He II, and C IV/He II (e.g., Feltre et al. 2016; Hirschmann et al. 2019, 2022). Unfortunately, many of these very high ionization lines are often weak and thus may not always have well-constrained ratios, as is the case for the object studied in this work. Photoionization modeling (e.g., from Cleri et al. 2023), along with comparisons to data across a broad redshift range ($0 \lesssim z \lesssim 8.5$), suggest that sources of very highly ionizing photons ($>54.42$ eV; Berg et al. 2021) may easily be confounded with AGNs by traditional BPT-style diagnostics. This result has been shown with well-studied $z \sim 0$ extreme metal-poor dwarf star-forming galaxies, which have been used as analogs to galaxies in the EoR (e.g., Berg et al. 2019, 2021; Olivier et al. 2022).

Many recent studies from early JWST data have shown exactly this: high-ionization, low-metallicity star formation may produce line ratios consistent with AGNs in the lower-redshift diagnostics (e.g., Brinchmann 2023; Katz et al. 2023; Trump et al. 2023; Trussler et al. 2022). From the other side, other work has shown known AGNs that produce line ratios consistent with star-forming galaxies (Übler et al. 2023). This growing body of work suggests that the dichotomous classification of a galaxy as either AGN-dominated or star-formation-dominated is inadequate to accurately describe a galaxy's ionizing spectrum in the early Universe.

These works suggest a larger "composite" region of BPT-style diagrams as a function of redshift. This allows for contributions to the ionizing spectrum from multiple sources (e.g., an accreting BH and extreme metal-poor star formation). As more exotic systems in the early Universe are found, evolved versions of these diagnostics and their use with other information about the galaxy (e.g., the analysis of broad Balmer lines and SED fits as shown in this work) will become increasingly necessary to discriminate between sources of ionization.

### 7.4. Implications for the Reionization of This Overdense Region at $z = 8.7$

The presence of an AGN in this object adds another layer of intrigue to this region. This object is one of two spectroscopically confirmed galaxies discussed by Larson et al. (2022), which reside within a larger photometric overdensity of five bright HST-selected galaxies discussed by Finkelstein et al. (2022a). The CEERS NIRCam data also show further evidence of an overabundance of $z \sim 8.5$–9 galaxies in this field, as shown in Finkelstein et al. (2022b) and studied in more detail by Larson et al. (2023, in preparation) and Whitler et al. (2023). Larson et al. (2022) discussed the ability of the galaxies in this region to ionize their surroundings, allowing Ly$\alpha$ to be visible. They found that an overdense region could provide the needed emissivity to reionize a large ($\gtrsim 1$ pMpc) bubble, necessary for Ly$\alpha$ to redshift out of resonance. While at the observed epoch, the AGN does not appear to dominate the rest-UV emission, potential past periods of super-Eddington growth could have emitted large amounts of ionizing photons, potentially ionizing a large volume around this region.

## 8. Conclusions

We present the discovery of an accreting SMBH at $z = 8.679$, using spectroscopy from NIRSpec and NIRCam/WFSS and imaging from NIRCam and MIRI from the JWST CEERS Survey (Bagley et al. 2023; Finkelstein et al. 2022b; S. L. Finkelstein et al. 2023, in preparation). This source, denoted here as CEERS_1019, was initially identified as a $z \sim 8$ photometric Ly$\alpha$-break dropout candidate by Roberts-Borsani et al. (2016b), with spectroscopic confirmation via Ly$\alpha$ emission using Keck/MOSFIRE by Zitrin et al. (2015). We detect several strong rest-UV and rest-optical emission lines using medium-resolution JWST/NIRSpec spectroscopy ($R \sim 1000$) covering 1–5 $\mu$m. From this work, we measure a significant broad component of the H$\beta$ emission line with FWHM $\sim 1200$ km s$^{-1}$, which we conclude originates from AGN activity.

Our measurements are based on observations from several JWST instruments, but our key results are derived from the 1–5 $\mu$m medium-resolution grating spectra from NIRSpec. We use an automated line-finding code (based on Larson et al. 2018) to identify significant emission features in an unbiased and systematic way, running line-injection simulations to estimate accurate line flux uncertainties. Due to the very high S/N of the [O III] 5008 Å emission line, most of our emission-line fits are tied to the redshift and FWHM of this line. We perform more customized fits when required, including for known doublets, as well as observed broad lines. We also note that a few NIRSpec-detected lines are also observable in the NIRCam WFSS spectra, albeit at lower S/N.

Our key result is that the observed H$\beta$ emission line has a clear broad component, comprising $\sim$half of the emission-line flux, with an FWHM $\sim 1200$ km s$^{-1}$. As this broad component is not seen in the stronger [O III] lines (as would be the case in large-scale outflows), we conclude that its origin is from a broadline region around an AGN. This is supported by weak N V, N IV, and Mg II emission, as well as weak broad C III] emission, and the morphology, which shows a compact point source among three Sérsic-like clumps.

We explore the properties of both this accreting SMBH as well as the host galaxy. Constraints from the continuum SED, including photometry from HST, as well as NIRCam and MIRI, show that the continuum emission is dominated by stellar light, particularly in the rest-UV, and that the stellar population is modestly massive (log $M/M_\odot \sim 9.5$) and heavily star-forming (log sSFR $\sim -7.9$).

From the width and flux of the broad H$\beta$ emission feature, we estimate the mass of the SMBH to be log ($M/M_\odot$) $= 6.95 \pm 0.37$ and that it is accreting at 1.2 ($\pm 0.5$) times the Eddington limit. From the ratios of narrow emission lines, we find that the gas in this galaxy is modestly metal-poor ($\sim 0.1$ $Z_\odot$) with little dust attenuation, dense, and highly ionized. Similar to other recent JWST results, CEERS_1019 sits elevated over most star-forming models in diagnostic line ratio diagrams. While this could be interpreted as evidence that the AGN significantly contributes to the narrow lines, the presence of other presumably star-forming-dominated galaxies in a similar line ratio regime means that we cannot rule out stellar-dominated ionization.

We discuss the implications of the presence of this BH early in cosmic history. We find that it is difficult to explain an SMBH of this mass at $z \sim 8.7$ with a stellar seed unless periodic episodes of super-Eddington accretion are possible. Alternatively, Eddington-limited accretion from a massive ($\sim 10^{4-6}$ $M_\odot$) DCBH seed could reach the target mass by the observed epoch. Either scenario is somewhat exotic—the uncovering of a larger population of early SMBHs will place further constraints on their seeding and growth mechanisms.

We conclude by noting that while the broad H$\beta$ component is statistically significant ($2.5\sigma$) and is significantly required by the fit ($\Delta_{\rm BIC} \sim 3$), we can only detect this component in one Balmer line. However, the next few months will see a MIRI spectrum obtained by the MIRI Guaranteed Time Observation (GTO) team (PID 1262), covering H$\alpha$. If, as we have concluded, a broadline AGN is present in this source, then the upcoming MIRI H$\alpha$ spectra should show a well-detected broad line.

## Acknowledgments

We sincerely thank all of the engineers, scientists, technicians, staff, other humans, and their families who spent decades of their lives making JWST possible, with a special thanks to those who have spent countless hours this past year commissioning and operating the telescope (Rigby et al. 2023) and providing calibration and pipeline updates. We also thank our other colleagues in the CEERS collaboration for their hard work and valuable contributions to this project.

We thank Xiaohui Fan, Dan Stark, and Rafaella Schneider for their helpful conversations. This work acknowledges support from the NASA/ESA/CSA JWST through the Space Telescope Science Institute (STScI), operated by the Association of Universities for Research in Astronomy, Incorporated, under NASA contract NAS5-03127. Support for program No. JWST-ERS01345 was provided through a grant from the STScI under NASA contract NAS5-03127.

R.L.L., S.L.F., and M.B. acknowledge that they work at an institution, the University of Texas at Austin, that sits on indigenous land. The Tonkawa lived in central Texas, and the Comanche and Apache moved through this area. We pay our respects to all the American Indian and Indigenous Peoples and communities who have been or have become a part of these lands and territories in Texas. We are grateful to be able to live, work, collaborate, and learn on this piece of Turtle Island.

The authors acknowledge the Texas Advanced Computing Center (TACC; tacc.utexas.edu) at the University of Texas at Austin for providing database and grid resources that have contributed to the research results reported within this paper. This work has used the Rainbow Cosmological Surveys Database, operated by the Centro de Astrobiologa (CAB), CSIC-INTA, partnered with the University of California Observatories at Santa Cruz (UCO/Lick, UCSC).

T.A.H. and A.Y. are supported by appointment to the NASA Postdoctoral Program (NPP) at NASA Goddard Space Flight Center, administered by Oak Ridge Associated Universities under contract with NASA. C.P. thanks Marsha and Ralph Schilling for the generous support of this research. This work benefited from support from the George P. and Cynthia Woods Mitchell Institute for Fundamental Physics and Astronomy at Texas A&M University. D.K. acknowledges support from NASA grants JWST-ERS-01345 and JWST-AR-02446. J.R.T. acknowledges support from NSF grant CAREER-1945546. D.B. and M.H.-C. thank the Programme National de Cosmologie et Galaxies and CNES for their support. R.A. acknowledges support from Fondecyt Regular 1202007. S.F. acknowledges funding from NASA through the NASA Hubble Fellowship grant HST-HF2-51505.001-A awarded by STScI. A.Z. acknowledges support from grant No. 2020750 from the United States–Israel Binational Science Foundation (BSF) and grant No. 2109066 from the United States National Science Foundation (NSF) and from the Ministry of Science & Technology, Israel.

Some of the data presented in this paper were obtained from the Mikulski Archive for Space Telescopes (MAST) at the Space Telescope Science Institute. The specific observations analyzed can be accessed via doi:10.17909/z7p0-8481 and at the CEERS Mast Archive Page.

*Facilities:* HST (ACS and WFC3), JWST (NIRCam, MIRI, NIRSpec, and NIRCam/WFSS), Spitzer (IRAC and MIPS), Keck (MOSFIRE), SCUBA-2, VLA, Chandra X-Ray Observatory, Herschel (PACS and SPIRE), and the Texas Advanced Computing Center (TACC).

*Software:* BPASS v2.0 and v2.2.1 (Eldridge et al. 2017; Stanway & Eldridge 2018), CLOUDY v17.0 (Ferland et al. 2017), EAZY (Brammer et al. 2008), Prospector (Johnson et al. 2021), Cigale (Boquien et al. 2019; Yang et al. 2020, 2022), astropy (The Astropy Collaboration et al. 2018), topcat (Taylor 2011), Galfit (Peng et al. 2002, 2010), statmorph (Rodriguez-Gomez et al. 2019), BAGPIPES (Carnall et al. 2018), MAPPINGS V (Sutherland et al. 2018; Kewley et al. 2019a), FAST v1.1 (Kriek et al. 2009; Aird et al. 2018), IDL Astronomy Library: idlastro.gsfc.nasa.gov (Landsman 1993), matplotlib (Hunter 2007), NumPy (Harris et al. 2020), photutils (Bradley et al. 2020), SourceExtractor (Bertin & Arnouts 1996), SciPy (Virtanen et al. 2020), STScI JWST Calibration Pipeline (jwst-pipeline.readthedocs.io; Rigby et al. 2023).

## Appendix A
## Far-infrared and Submillimeter Constraints

Recent studies have shown that a significant fraction of high-redshift AGNs exhibit large rest-frame infrared luminosities, with values ranging from $10^{12}$ to $10^{13}$ $L_\odot$ (e.g., Decarli et al. 2018). Here we report on the far-infrared and submillimeter photometric constraints on this source, exploiting the deep ancillary data in the field. As shown in Figure 12, a $\sim 6\sigma$ SCUBA-2 detection (yellow contours) with a flux of $S_{850\mu m} = 1.9 \pm 0.3$ mJy (Zavala et al. 2017) is found 2″ away from the position of CEERS_1019 (red circle). Zavala et al. (2018) associated this emission as originating from a nearby $z_{phot} \approx 3$ galaxy (named 850.44) marked by a blue cross in Figure 12. This nearby source is also detected with Spitzer at 24 $\mu$m and at 100 $\mu$m with Herschel, as shown in panels 2 and 3 of Figure 12. A recent survey was conducted with the JVLA at 3 GHz (PI: M. Dickinson; see also Jimenez-Andrade et al. 2023, in preparation), which also found detectable emission at this location, likely associated with the nearby source (the right panel in Figure 12).

If this submillimeter detection was instead emitting from CEERS_1019 at $z = 8.679$, it would imply an infrared luminosity of $L_{IR} \sim 3 \times 10^{12} L_\odot$. Although this is in line with the luminosity of $z \sim 7$ quasars detected with the Atacama Large Millimeter/submillimeter Array (e.g., Decarli et al. 2018), it would exceed the Eddington limit of our target by a factor of $\sim 10$. This supports the assumption that the submillimeter emission arises (mainly) from a different neighboring source.

## Appendix B
## X-Ray Constraints

The Chandra X-ray Observatory took an 800 ks exposure over the EGS field (Nandra et al. 2015), but there is no emission detected at the location of CEERS_1019 in these data (see Figure 13). Adopting a 0.5–10 keV sensitivity of $8.22 \times 10^{-16}$ erg cm$^{-1}$ s$^{-1}$ (Nandra et al. 2015) and assuming a photon index ($\Gamma$) of 1.4, we estimate an upper limit of $L_X < 10^{44.2}$ erg s$^{-1}$. This constraint places CEERS_1019 around or below the knee luminosity ($L_X^*$) of the AGN X-ray luminosity function (i.e., the Seyfert regime) at lower redshifts ($z \approx 0$–5; e.g., Aird et al. 2015). Thanks to its unprecedented sensitivity, JWST begins to uncover the Seyfert-like AGN population in the early universe.

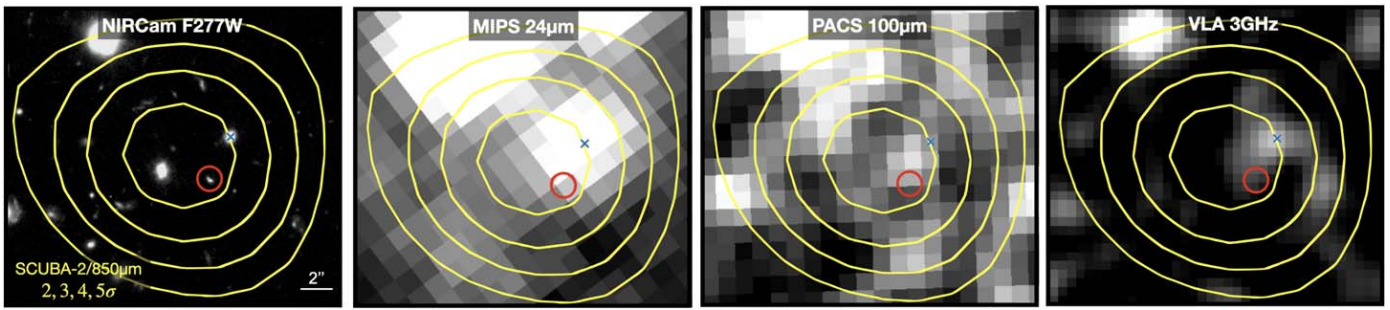

**Figure 12.** From left to right: JWST/NIRCam F277W; Spitzer/MIPS 24 $\mu$m; Herschel/PACS 100 $\mu$m; and VLA 3 GHz (north is up, east is to the left). All the 20″ × 20″ cutouts are centered on the coordinates of the 6$\sigma$ SCUBA-2/850 $\mu$m detection (yellow contours) found around the location of CEERS_1019 (red circle). The blue cross marks the position of a $z \sim 3$ galaxy detected at 24 $\mu$m and 3 GHz, which is likely the correct counterpart and the main contributor to the SCUBA-2 detection.

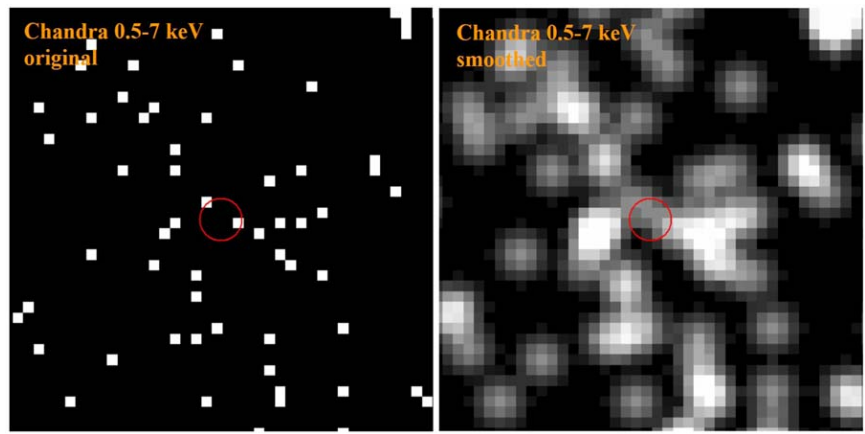

**Figure 13.** Full-band (0.5–7 keV) X-ray images from Nandra et al. (2015). The format is similar to Figure 12. The left panel shows the original image, where each white pixel indicates one (or multiple) X-ray photon(s). The right panel is a smoothed version of the left. Noise dominates in the region around the source.

## Appendix C
## NIRSpec Prism Spectrum

The NIRSpec/PRISM observation of CEERS_1019 (see Figure 14) was not included in the analysis of this source, as it was observed in 2022 December and was contaminated by a short. When these CEERS observations were rescheduled for 2023 February, the telescope had rotated such that the MSA had to be redesigned, and this source was no longer included. We include the spectrum here for completeness and to illustrate that several lines are still visible in the spectrum, including N IV], C III], H$\beta$, and [O III], despite the contamination. Recent publications, including Heintz et al. (2022; where they identify the source as CEERS-z8684) and Harikane et al. (2023), have shown this PRISM spectrum and also detected these lines.

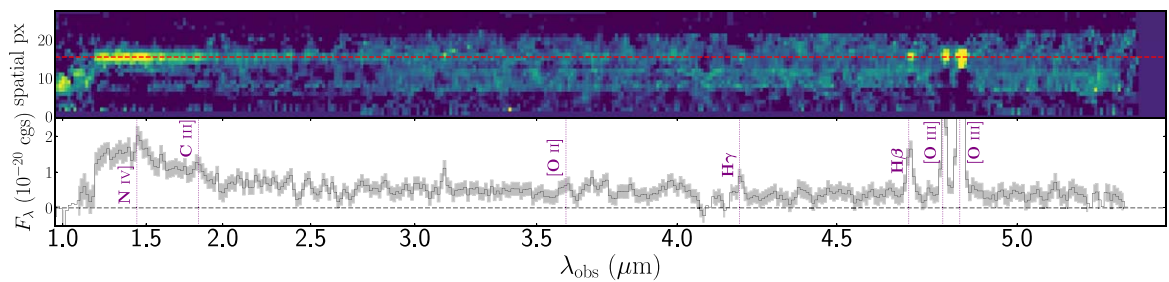

**Figure 14.** 2D and 1D spectra of the NIRSpec/PRISM spectrum of CEERS_1019, taken in 2022 December, which encountered a short contaminating a majority of the data. We show the spectra here for completeness and also to illustrate that many of the emission lines are visible, including N IV], C III], H$\beta$, and [O III], even though the data are contaminated.

## ORCID iDs

Rebecca L. Larson https://orcid.org/0000-0003-2366-8858
Steven L. Finkelstein https://orcid.org/0000-0001-8519-1130
Dale D. Kocevski https://orcid.org/0000-0002-8360-3880
Taylor A. Hutchison https://orcid.org/0000-0001-6251-4988
Jonathan R. Trump https://orcid.org/0000-0002-1410-0470
Pablo Arrabal Haro https://orcid.org/0000-0002-7959-8783
Volker Bromm https://orcid.org/0000-0003-0212-2979
Nikko J. Cleri https://orcid.org/0000-0001-7151-009X
Mark Dickinson https://orcid.org/0000-0001-5414-5131
Seiji Fujimoto https://orcid.org/0000-0001-7201-5066
Jeyhan S. Kartaltepe https://orcid.org/0000-0001-9187-3605
Anton M. Koekemoer https://orcid.org/0000-0002-6610-2048
Casey Papovich https://orcid.org/0000-0001-7503-8482
Nor Pirzkal https://orcid.org/0000-0003-3382-5941
Sandro Tacchella https://orcid.org/0000-0002-8224-4505
Jorge A. Zavala https://orcid.org/0000-0002-7051-1100
Micaela Bagley https://orcid.org/0000-0002-9921-9218
Peter Behroozi https://orcid.org/0000-0002-2517-6446
Jaclyn B. Champagne https://orcid.org/0000-0002-6184-9097
Justin W. Cole https://orcid.org/0000-0002-6348-1900
Intae Jung https://orcid.org/0000-0003-1187-4240
Alexa M. Morales https://orcid.org/0000-0003-4965-0402
Guang Yang https://orcid.org/0000-0001-8835-7722
Haowen Zhang https://orcid.org/0000-0002-4321-3538
Adi Zitrin https://orcid.org/0000-0002-0350-4488
Ricardo O. Amorín https://orcid.org/0000-0001-5758-1000
Denis Burgarella https://orcid.org/0000-0002-4193-2539
Caitlin M. Casey https://orcid.org/0000-0002-0930-6466
Óscar A. Chávez Ortiz https://orcid.org/0000-0003-2332-5505
Isabella G. Cox https://orcid.org/0000-0002-1803-794X
Katherine Chworowsky https://orcid.org/0000-0003-4922-0613
Adriano Fontana https://orcid.org/0000-0003-3820-2823
Eric Gawiser https://orcid.org/0000-0003-1530-8713
Andrea Grazian https://orcid.org/0000-0002-5688-0663
Norman A. Grogin https://orcid.org/0000-0001-9440-8872
Santosh Harish https://orcid.org/0000-0003-0129-2079
Nimish P. Hathi https://orcid.org/0000-0001-6145-5090
Michaela Hirschmann https://orcid.org/0000-0002-3301-3321
Benne W. Holwerda https://orcid.org/0000-0002-4884-6756
Stéphanie Juneau https://orcid.org/0000-0002-0000-2394
Gene C. K. Leung https://orcid.org/0000-0002-9393-6507
Ray A. Lucas https://orcid.org/0000-0003-1581-7825
Elizabeth J. McGrath https://orcid.org/0000-0001-8688-2443
Pablo G. Pérez-González https://orcid.org/0000-0003-4528-5639
Jane R. Rigby https://orcid.org/0000-0002-7627-6551
Lise-Marie Seillé https://orcid.org/0000-0001-7755-4755
Raymond C. Simons https://orcid.org/0000-0002-6386-7299
Alexander de la Vega https://orcid.org/0000-0002-6219-5558
Benjamin J. Weiner https://orcid.org/0000-0001-6065-7483
Stephen M. Wilkins https://orcid.org/0000-0003-3903-6935
L. Y. Aaron Yung https://orcid.org/0000-0003-3466-035X

## References

Aird, J., Coil, A. L., & Georgakakis, A. 2018, MNRAS, 474, 1225
Aird, J., Coil, A. L., Georgakakis, A., et al. 2015, MNRAS, 451, 1892
Amorín, R., Pérez-Montero, E., Vílchez, J. M., & Papaderos, P. 2012, ApJ, 749, 185
Amorín, R., Pérez-Montero, E., Contini, T., et al. 2015, A&A, 578, A105
Arrabal Haro, P., Dickinson, M., Finkelstein, S. L., et al. 2023, ApJL, 951, L22
Asplund, M., Amarsi, A. M., & Grevesse, N. 2021, A&A, 653, A141
Bañados, E., Venemans, B. P., Mazzucchelli, C., et al. 2018, Natur, 553, 473
Backhaus, B. E., Trump, J. R., Cleri, N. J., et al. 2022, ApJ, 926, 161
Bagley, M. B., Finkelstein, S. L., Koekemoer, A. M., et al. 2023, ApJL, 946, L12
Baldwin, J. A., Phillips, M. M., & Terlevich, R. 1981, PASP, 93, 5
Beers, T. C., Flynn, K., & Gebhardt, K. 1990, AJ, 100, 32
Beichman, C. A., Rieke, M., Eisenstein, D., et al. 2012, Proc. SPIE, 8442, 84422N
Berg, D. A., Chisholm, J., Erb, D. K., et al. 2019, ApJL, 878, L3
Berg, D. A., Chisholm, J., Erb, D. K., et al. 2021, ApJ, 922, 170
Bertin, E., & Arnouts, S. 1996, A&AS, 117, 393
Böker, T., Beck, T. L., Birkmann, S. M., et al. 2023, PASP, 135, 8001
Boquien, M., Burgarella, D., Roehlly, Y., et al. 2019, A&A, 622, A103
BradleyL., SipőczB., RobitailleT., et al. (2020) astropy/photutils: v1.0.1, Zenodo, doi:10.5281/zenodo.4049061
Brammer, G. B., van Dokkum, P. G., & Coppi, P. 2008, ApJ, 686, 1503
Brinchmann, J. 2023, MNRAS, Advance Access
Bromm, V., & Loeb, A. 2003, ApJ, 596, 34
Bunker, A. J., Saxena, A., Cameron, A. J., et al. 2023, arXiv:2302.07256
BushouseH., EisenhamerJ., DenchevaN., et al. (2022a) JWST Calibration Pipelinev1.8.5 Zenodo, doi:10.5281/zenodo.7429939
BushouseH., EisenhamerJ., DenchevaN., et al. (2022b) JWST Calibration Pipeline1.7.2 Zenodo, doi:10.5281/zenodo.7071140
Cackett, E. M., Bentz, M. C., & Kara, E. 2021, iSci, 24, 102557
Calzetti, D. 2001, PASP, 113, 1449
Calzetti, D., Kinney, A. L., & Storchi-Bergmann, T. 1994, ApJ, 429, 582
Cameron, A. J., Katz, H., Rey, M. P., & Saxena, A. 2023, MNRAS, 523, 3516
Carnall, A. C., McLure, R. J., Dunlop, J. S., & Davé, R. 2018, MNRAS, 480, 4379
Cleri, N. J., Yang, G., Papovich, C., et al. 2023, ApJ, 948, 112
Cleri, N. J., Olivier, G. M., Hutchison, T. A., et al. 2023, ApJ, 953, 10
Coil, A. L., Aird, J., Reddy, N., et al. 2015, ApJ, 801, 35
Dayal, P., Volonteri, M., Choudhury, T. R., et al. 2020, MNRAS, 495, 3065
Decarli, R., Walter, F., Venemans, B. P., et al. 2018, ApJ, 854, 97
Dekel, A., Sarkar, K. S., Birnboim, Y., Mandelker, N., & Li, Z. 2023, MNRAS, 523, 3201
Dijkstra, M., Haiman, Z., Rees, M. J., & Weinberg, D. H. 2004, ApJ, 601, 666
Dong, X., Wang, T., Wang, J., et al. 2008, MNRAS, 383, 581
Eldridge, J. J., Stanway, E. R., Xiao, L., et al. 2017, PASA, 34, e058
Endsley, R., Stark, D. P., Bouwens, R. J., et al. 2022, MNRAS, 517, 5642
Erb, D. K., Pettini, M., Shapley, A. E., et al. 2010, ApJ, 719, 1168
Fan, X., Banados, E., & Simcoe, R. A. 2022, arXiv:2212.06907
Fan, X., Strauss, M. A., Becker, R. H., et al. 2006, AJ, 132, 117
Farina, E. P., Schindler, J.-T., Walter, F., et al. 2022, ApJ, 941, 106
Feltre, A., Charlot, S., & Gutkin, J. 2016, MNRAS, 456, 3354
Ferland, G. J., Chatzikos, M., Guzmán, F., et al. 2017, RMxAA, 53, 385
Ferruit, P., Jakobsen, P., Giardino, G., et al. 2022, A&A, 661, A81
Finkelstein, S. L., D'Aloisio, A., Paardekooper, J.-P., et al. 2019, ApJ, 879, 36
Finkelstein, S. L., Bagley, M., Song, M., et al. 2022a, ApJ, 928, 52
Finkelstein, S. L., Bagley, M. B., Arrabal Haro, P., et al. 2022b, ApJL, 940, 55
Fontanot, F., Cristiani, S., Grazian, A., et al. 2023, MNRAS, 520, 740
Foreman-Mackey, D., Hogg, D. W., Lang, D., & Goodman, J. 2013, PASP, 125, 306
Freese, K., Rindler-Daller, T., Spolyar, D., & Valluri, M. 2016, RPPh, 79, 066902
Fujimoto, S., Brammer, G. B., Watson, D., et al. 2022, Natur, 604, 261
Fujimoto, S., Arrabal Haro, P., Dickinson, M., et al. 2023, ApJL, 949, 25
Furtak, L. J., Zitrin, A., Plat, A., et al. 2023, ApJ, 952, 142
Gardner, J. P., Mather, J. C., Abbott, R., et al. 2023, PASP, 135, 8001
Giallongo, E., Grazian, A., Fiore, F., et al. 2019, ApJ, 884, 19

Gibson, R. R., Jiang, L., Brandt, W. N., et al. 2009, ApJ, 692, 758
Glikman, E., Rusu, C. E., Chen, G. C. F., et al. 2023, ApJ, 943, 25
Goodman, J., & Weare, J. 2010, Comm App Math Comp Sci, 5, 65
Grazian, A., Giallongo, E., Fiore, F., et al. 2020, ApJ, 897, 94
Grazian, A., Giallongo, E., Boutsia, K., et al. 2022, ApJ, 924, 62
Greene, J. E., & Ho, L. C. 2005, ApJ, 630, 122
Greene, J. E., Alexandroff, R., Strauss, M. A., et al. 2014, ApJ, 788, 91
Greene, J. E., Seth, A., Kim, M., et al. 2016, ApJL, 826, L32
Greene, T. P., Kelly, D. M., Stansberry, J., et al. 2017, JATIS, 3, 035001
Grogin, N. A., Kocevski, D. D., Faber, S. M., et al. 2011, ApJS, 197, 35
Habouzit, M., Li, Y., Somerville, R. S., et al. 2021, MNRAS, 503, 1940
Haiman, Z. 2004, ApJ, 613, 36
Hamann, F., & Ferland, G. 1999, ARA&A, 37, 487
Harikane, Y., Nakajima, K., Ouchi, M., et al. 2023, arXiv:2304.06658
Harris, C. R., Millman, K. J., van der Walt, S. J., et al. 2020, Natur, 585, 357
Heintz, K. E., Brammer, G. B., Giménez-Arteaga, C., et al. 2022, arXiv:2212.02890
Hirschmann, M., Charlot, S., Feltre, A., et al. 2019, MNRAS, 487, 333
Hirschmann, M., Charlot, S., Feltre, A., et al. 2022, arXiv:2212.02522
Hogarth, L., Amorín, R., Vílchez, J. M., et al. 2020, MNRAS, 494, 3541
Horne, K. 1986, PASP, 98, 609
Hunter, J. D. 2007, CSE, 9, 90
Ilie, C., Freese, K., Valluri, M., Iliev, I. T., & Shapiro, P. R. 2012, MNRAS, 422, 2164
Inayoshi, K., Haiman, Z., & Ostriker, J. P. 2016, MNRAS, 459, 3738
Inayoshi, K., Visbal, E., & Haiman, Z. 2020, ARA&A, 58, 27
Iyer, K. G., Gawiser, E., Faber, S. M., et al. 2019, ApJ, 879, 116
Izotov, Y. I., Schaerer, D., Worseck, G., et al. 2018, MNRAS, 474, 4514
Izumi, T., Matsuoka, Y., Fujimoto, S., et al. 2021, ApJ, 914, 36
Jakobsen, P., Ferruit, P., Alves de Oliveira, C., et al. 2022, A&A, 661, A80
Jeffreys, H. 1961, The Theory of Probability (3rd edn.; Oxford Univ. Press)
Jeon, M., Pawlik, A. H., Bromm, V., & Milosavljević, M. 2014, MNRAS, 440, 3778
Jeon, M., Pawlik, A. H., Greif, T. H., et al. 2012, ApJ, 754, 34
Johnson, B. D., Leja, J., Conroy, C., & Speagle, J. S. 2021, ApJS, 254, 22
Johnson, J. L., & Bromm, V. 2007, MNRAS, 374, 1557
Juneau, S., Dickinson, M., Alexander, D. M., & Salim, S. 2011, ApJ, 736, 104
Juneau, S., Bournaud, F., Charlot, S., et al. 2014, ApJ, 788, 88
Jung, I., Finkelstein, S. L., Dickinson, M., et al. 2020, ApJ, 904, 144
Kaspi, S., Smith, P. S., Netzer, H., et al. 2000, ApJ, 533, 631
Katz, H., Saxena, A., Cameron, A. J., et al. 2023, MNRAS, 518, 592
Keenan, F. P., Feibelman, W. A., & Berrington, K. A. 1992, ApJ, 389, 443
Kennicutt, R. C., & Evans, N. J. 2012, ARA&A, 50, 531
Kewley, L. J., Nicholls, D. C., Sutherland, R., et al. 2019a, ApJ, 880, 16
Kewley, L. J., Nicholls, D. C., & Sutherland, R. S. 2019b, ARA&A, 57, 511
Kobulnicky, H. A., & Kewley, L. J. 2004, ApJ, 617, 240
Kocevski, D. D., Onoue, M., Inayoshi, K., et al. 2023, arXiv:2302.00012
Koekemoer, A. M., Faber, S. M., Ferguson, H. C., et al. 2011, ApJS, 197, 36
Kormendy, J., & Ho, L. C. 2013, ARA&A, 51, 511
Kriek, M., van Dokkum, P. G., Franx, M., Illingworth, G. D., & Magee, D. K. 2009, ApJL, 705, L71
Kroupa, P., Subr, L., Jerabkova, T., & Wang, L. 2020, MNRAS, 498, 5652
Landsman, W. B. 1993, in ASP Conf. Ser. 52, Astronomical Data Analysis Software and Systems II, ed. R. J. Hanisch, R. J. V. Brissenden, & J. Barnes (San Francisco, CA: ASP), 246
Larson, R. L., Finkelstein, S. L., Pirzkal, N., et al. 2018, ApJ, 858, 94
Larson, R. L., Finkelstein, S. L., Hutchison, T. A., et al. 2022, ApJ, 930, 104
Latif, M. A., Khochfar, S., Schleicher, D., & Whalen, D. J. 2021, MNRAS, 508, 1756
Lauer, T. R., Tremaine, S., Richstone, D., & Faber, S. M. 2007, ApJ, 670, 249
Leitherer, C., Ekström, S., Meynet, G., et al. 2014, ApJS, 212, 14
Liddle, A. R. 2004, MNRAS, 351, L49
MacLeod, C. L., Ivezić, Ž., Sesar, B., et al. 2012, ApJ, 753, 106
Madau, P., Haardt, F., & Dotti, M. 2014, ApJL, 784, L38
Mainali, R., Zitrin, A., Stark, D. P., et al. 2018, MNRAS, 479, 1180
Markwardt, C. B. 2009, in ASP Conf. Ser. 411, Astronomical Data Analysis Software and Systems XVIII, ed. D. A. Bohlender, D. Durand, & P. Dowler (San Francisco, CA: ASP), 251
Mason, C. A., & Gronke, M. 2020, MNRAS, 499, 1395
Massonneau, W., Volonteri, M., Dubois, Y., & Beckmann, R. S. 2023, A&A, 670, A180
Matsuoka, K., Nagao, T., Maiolino, R., et al. 2017, A&A, 608, A90
Mortlock, D. J., Warren, S. J., Venemans, B. P., et al. 2011, Natur, 474, 616
Nandra, K., Laird, E. S., Aird, J. A., et al. 2015, ApJS, 220, 10
Natarajan, P., Pacucci, F., Ferrara, A., et al. 2017, ApJ, 838, 117
Netzer, H. 2015, ARA&A, 53, 365
Nicholls, D. C., Kewley, L. J., & Sutherland, R. S. 2020, PASP, 132, 033001
Nicholls, D. C., Sutherland, R. S., Dopita, M. A., Kewley, L. J., & Groves, B. A. 2017, MNRAS, 466, 4403
Olivier, G. M., Berg, D. A., Chisholm, J., et al. 2022, ApJ, 938, 16
Osterbrock, D. E. 1989, Astrophysics of Gaseous Nebulae and Active Galactic Nuclei (Mill Valley, CA: Univ. Science Books)
Pacucci, F., & Loeb, A. 2022, ApJL, 940, L33
Papovich, C., Simons, R. C., Estrada-Carpenter, V., et al. 2022, ApJ, 937, 22
Papovich, C., Cole, J., Yang, G., et al. 2023, ApJL, 949, L18
Park, H., Jung, I., Song, H., et al. 2021, ApJ, 922, 263
Peng, C. Y., Ho, L. C., Impey, C. D., & Rix, H.-W. 2002, AJ, 124, 266
Peng, C. Y., Ho, L. C., Impey, C. D., & Rix, H.-W. 2010, AJ, 139, 2097
Pérez-Montero, E., Amorín, R., Sánchez Almeida, J., et al. 2021, MNRAS, 504, 1237
Pezzulli, E., Valiante, R., & Schneider, R. 2016, MNRAS, 458, 3047
Planck Collaboration, Aghanim, N., Akrami, Y., et al. 2020, A&A, 641, A6
Polletta, M. 2008, A&A, 480, L41
Polletta, M., Tajer, M., Maraschi, L., et al. 2007, ApJ, 663, 81
Regan, J. A., Downes, T. P., Volonteri, M., et al. 2019, MNRAS, 486, 3892
Reines, A. E., & Volonteri, M. 2015, ApJ, 813, 82
Ricarte, A., & Natarajan, P. 2018a, MNRAS, 481, 3278
Ricarte, A., & Natarajan, P. 2018b, MNRAS, 474, 1995
Richards, G. T., Strauss, M. A., Fan, X., et al. 2006, AJ, 131, 2766
Rieke, M. J., Kelly, D., & Horner, S. 2005, Proc. SPIE, 5904, 1
Rieke, M. J., Baum, S. A., Beichman, C. A., et al. 2003, Proc. SPIE, 4850, 478
Rieke, M. J., Kelly, D. M., Misselt, K., et al. 2023, PASP, 135, 028001
Rigby, J., Perrin, M., McElwain, M., et al. 2023, PASP, 135, 8001
Roberts-Borsani, G. W., Bouwens, R. J., Oesch, P. A., et al. 2016a, ApJ, 823, 143
Roberts-Borsani, G. W., Bouwens, R. J., Oesch, P. A., et al. 2016b, ApJ, 823, 143
Rodriguez-Gomez, V., Snyder, G. F., Lotz, J. M., et al. 2019, MNRAS, 483, 4140
Salpeter, E. E. 1955, ApJ, 121, 161
Sanders, R. L., Shapley, A. E., Jones, T., et al. 2023, ApJ, 942, 24
Schaerer, D., Izotov, Y. I., Worseck, G., et al. 2022, A&A, 658, L11
Schartmann, M., Meisenheimer, K., Camenzind, M., Wolf, S., & Henning, T. 2005, A&A, 437, 861
Schmidt, E. O., Ferreiro, D., Vega Neme, L., & Oio, G. A. 2016, A&A, 596, A95
Senchyna, P., Plat, A., Stark, D. P., & Rudie, G. C. 2023, arXiv:2303.04179
Shapley, A. E., Steidel, C. C., Pettini, M., & Adelberger, K. L. 2003, ApJ, 588, 65
Silk, J. 2005, MNRAS, 364, 1337
Silva, L., Maiolino, R., & Granato, G. L. 2004, MNRAS, 355, 973
Smith, A., & Bromm, V. 2019, ConPh, 60, 111
Stanway, E. R., & Eldridge, J. J. 2018, MNRAS, 479, 75
Stern, J., & Laor, A. 2012, MNRAS, 423, 600
Storey, P. J., & Zeippen, C. J. 2000, MNRAS, 312, 813
Sutherland R., Dopita M., Binette L., & Groves B. (2018) MAPPINGS V: Astrophysical plasma modeling code, Astrophysics Source Code Library record ascl:1807.005
Tacchella, S., Finkelstein, S. L., Bagley, M., et al. 2022, ApJ, 927, 170
Tang, M., Stark, D. P., Chen, Z., et al. 2023, arXiv:2301.07072
Taylor M. (2011) TOPCAT: Tool for OPerations on Catalogues And Tables, Astrophysics Source Code Library record ascl:1101.010
The Astropy Collaboration, Price-Whelan, A. M., Sipőcz, B. M., et al. 2018, AJ, 156, 123
Trinca, A., Schneider, R., Maiolino, R., et al. 2023, MNRAS, 519, 4753
Trump, J. R., Sun, M., Zeimann, G. R., et al. 2015, ApJ, 811, 26
Trump, J. R., Arrabal Haro, P., Simons, R. C., et al. 2023, ApJ, 945, 35
Trussler, J. A. A., Conselice, C. J., Adams, N. J., et al. 2022, arXiv:2211.02038
Übler, H., Maiolino, R., Curtis-Lake, E., et al. 2023, arXiv:2302.06647
Vanden Berk, D. E., Richards, G. T., Bauer, A., et al. 2001, AJ, 122, 549
Veilleux, S., & Osterbrock, D. E. 1987, ApJS, 63, 295
Virtanen, P., Gommers, R., Oliphant, T. E., et al. 2020, NatMe, 17, 261
Volonteri, M., Habouzit, M., & Colpi, M. 2021, NatRP, 3, 732
Volonteri, M., Habouzit, M., & Colpi, M. 2023, MNRAS, 521, 241
Volonteri, M., & Rees, M. J. 2005, ApJ, 633, 624
Wang, F., Yang, J., Fan, X., et al. 2021, ApJL, 907, L1
Whitler, L., Stark, D. P., Endsley, R., et al. 2023, MNRAS, submitted
Williams, H., Kelly, P. L., Chen, W., et al. 2023, Sci, 380, 416
Woods, T. E., Agarwal, B., Bromm, V., et al. 2019, PASA, 36, e027
Wright, G. S., Rieke, G. H., Glasse, A., et al. 2023, PASP, 135, 048003
Yang, G., Boquien, M., Buat, V., et al. 2020, MNRAS, 491, 740

Yang, G., Papovich, C., Bagley, M. B., et al. 2021, ApJ, 908, 144
Yang, G., Boquien, M., Brandt, W. N., et al. 2022, ApJ, 927, 192
Yung, L. Y. A., Somerville, R. S., Finkelstein, S. L., et al. 2021, MNRAS, 508, 2706
Yung, L. Y. A., Somerville, R. S., Finkelstein, S. L., et al. 2023, MNRAS, 519, 1578
Zavala, J. A., Aretxaga, I., Geach, J. E., et al. 2017, MNRAS, 464, 3369
Zavala, J. A., Aretxaga, I., Dunlop, J. S., et al. 2018, MNRAS, 475, 5585
Zhang, H., Behroozi, P., Volonteri, M., et al. 2023a, MNRAS, 518, 2123
Zhang, H., Behroozi, P., Volonteri, M., et al. 2023b, MNRAS, 523, 69
Zitrin, A., Labbé, I., Belli, S., et al. 2015, ApJL, 810, L12